\pdfoutput=1
\documentclass[aps,pre,twocolumn,groupedaddress]{revtex4-1}
\usepackage{graphicx}
\usepackage{lipsum}  
\usepackage[utf8]{inputenc}
\usepackage{mathtools}

\usepackage{float}

\usepackage[margin=1in]{geometry} 
\usepackage[usenames,dvipsnames]{xcolor}
\usepackage{parskip} 

\usepackage{relsize}
\usepackage{amsmath}    
\usepackage{amssymb}
\usepackage{bm}

\usepackage{hyperref}
\usepackage{latexsym}
\usepackage{verbatim}
\usepackage{color}
\usepackage{xcolor}
\enlargethispage{\baselineskip}

\def\beq{\begin{equation}}
\def\eeq{\end{equation}}

\def\beq{\begin{equation}}                          
\def\eeq{\end{equation}}                          
\def\bea{\begin{eqnarray}}                          
\def\eea{\end{eqnarray}}

\DeclareRobustCommand{\uvec}[1]{{%
  \ifcsname uvec#1\endcsname
     \csname uvec#1\endcsname
   \else
    \bm{\hat{\mathbf{#1}}}%
   \fi
}}

\preprint{}

\begin{document}

\title{Emergent Rotation of Passive Clusters in a Chiral Active Bath}
\author{Divya Kushwaha}
\email{Corresponding author: divyakushwaha.rs.phy22@itbhu.ac.in}
\affiliation{Indian Institute of Technology (BHU) Varanasi, India 221005}
\author{Abhra Puitandy}
\email{abhrapuitandy.rs.phy22@itbhu.ac.in}
\affiliation{Indian Institute of Technology (BHU) Varanasi, India 221005}
\author{Shradha Mishra}
\email[]{smishra.phy@itbhu.ac.in}
\affiliation{Indian Institute of Technology (BHU) Varanasi, India 221005}
\date{\today}
\begin{abstract}
{We investigate the dynamics of passive particles immersed in a bath of chiral active particles, focusing on the emergence of collective rotational motion. Using numerical simulations, we show that passive particles aggregate into clusters that can exhibit persistent rotation within a well-defined regime of size ratio and active particle packing fraction. This rotational state is characterized by the coexistence of internal structural order, enhanced shape fluctuations, and a coherent net torque generated by the surrounding active bath. Outside this regime, the dynamics remain predominantly diffusive, highlighting that sustained rotation is not ubiquitous but arises from a delicate interplay between geometry, activity, and chirality. Furthermore, we demonstrate that chirality heterogeneity disrupts rotational coherence, while a uniform chiral bath promotes strongly superdiffusive angular dynamics. These results provide new insights into the role of chirality and collective interactions in shaping the emergent behavior of active-passive mixtures.}
\end{abstract}
\maketitle
\section{Introduction}
Active matter systems consist of self-propelled particles that convert internal or environmental energy into directed motion, thereby breaking detailed balance at the single-particle level. They appear across many length scales, from cytoskeletal assemblies \cite{needleman2017active} and bacterial colonies \cite{dell2018growing} to animal groups \cite{hueschen2023wildebeest} and human crowds. These systems display a rich variety of collective behaviors \cite{bottinelli2016emergent,stenhammar2014phase,fily2012athermal} such as swarming \cite{nourhani2025biomimetic,liebchen2017collective}, clustering \cite{fehlinger2023collective}, and motility-induced phase separation \cite{cates2015motility,suma2014motility}. These emergent states make active matter a useful framework for studying how local non-equilibrium driving produces collective organization\cite{liebchen2017collective}.\\
A particularly fascinating class of active matter is formed by chiral active particles, which combine self-propulsion with intrinsic rotational motion arising from internal asymmetries or external torques \cite{liebchen2022chiral,puitandy2026spontaneous}. Chiral motion is commonly observed in biological microswimmers, including sperm cells and several bacterial species \cite{keaveny2009hydrodynamic,keaveny2013optimization,su2013sperm,jennings1901significance}, as well as in synthetic active colloids. The intrinsic chirality of these particles causes their trajectories to deviate from straight paths and follow circular or curved orbits, giving rise to a broad range of nonequilibrium phenomena such as hyperuniform states \cite{zhang2022hyperuniform}, self-sustained vortices \cite{caprini2024self,eswaran2024synchronized}, caging in active glasses \cite{mandal2016active}, and segregation in multicomponent mixtures \cite{kushwaha2025flocking}.\\
Concurrently, increasing interest has focused on mixtures of active and passive particles, which provide a platform for studying nonequilibrium self-organization, transport, and collective assembly in complex environments \cite{bechinger2016active,fernandez2025dynamics,serna2025sorting}. Experimental studies have shown that active baths modify the dynamics of passive colloids, leading to enhanced diffusion, clustering, segregation, and dynamic self-assembled structures \cite{koumakis2013targeted,palacci2013living,aubret2018targeted}. Complementary numerical investigations of dry active matter have demonstrated that active particles can generate effective interactions between passive inclusions, producing aggregation, phase separation, and collective behavior \cite{kushwaha2023phase,singh2022effective,stenhammar2015activity,ni2015tunable,dolai2018phase,gokhale2022dynamic,krafnick2015impact,yeo2016dynamics,deb2026hydrodynamic}. In wet active systems, hydrodynamic interactions further influence collective transport, active-mediated assembly, and gel-like structures of passive objects immersed in swimmer suspensions \cite{grober2023unconventional,yuan2024colloid,kushwaha2024percolation}.\\
A mechanism underlying passive-particle aggregation in active environments is the emergence of effective nonequilibrium attractions. Unlike equilibrium depletion interactions, these active-mediated forces resemble Casimir-like interactions that depend on particle activity, geometry, and confinement \cite{angelani2011effective,smrek2017small,dolai2018phase,ni2015tunable,ray2014casimir,liu2020constraint}.  
Introducing chirality into active Brownian particles generally suppresses their effective diffusivity and weakens the clustering tendency of passive colloids \cite{zaeifi2017effective,bickmann2022analytical,torrik2021dimeric, dabra2025depletion,reichhardt2019reversibility}, highlighting the role of rotational dynamics in activity-mediated assembly.\\
Beyond aggregation, active matter also generates rotational motion \cite{hargus2025odd} and chirality-induced self-assembled structures. Experiments have shown that active particles can assemble into self-spinning microstructures, demonstrating how microscopic activity can be converted into collective rotation at mesoscopic scales \cite{aubret2018targeted}. Chiral active systems exhibit phenomena including vortex formation, synchronization, rotating assemblies, and self-sustained rotational states \cite{caprini2024self,eswaran2024synchronized,liebchen2022chiral}. Recent studies have further shown that chirality heterogeneity can generate frustration, slow relaxation, and glassy dynamics, emphasizing the role of microscopic disorder in emergent collective behavior \cite{debets2023glassy}.
Despite these advances, most studies have largely focused either on active-mediated aggregation or on collective rotational phenomena separately. Consequently, the interplay between these two effects remains unclear. In particular, it is not known under what conditions passive aggregates formed through active-mediated interactions develop sustained rotational motion in a chiral active bath, or how this behavior is influenced by chirality heterogeneity. Addressing these questions provides insight into active self-assembly, transport, and collective dynamics in biological, soft-matter, and synthetic active systems.\\
In this work, we study the dynamics of passive particles immersed in a chiral active bath. All particles interact through soft repulsive forces, while a weak, short-range attraction between passive particles is introduced to maintain cluster cohesion. Experimentally, such cohesive interactions can be realized through depletion forces generated by non-adsorbing polymers, whose concentration and size control the strength and range of the effective attraction \cite{semwal2022tunable}. This effective short-range attraction provides a simplified description of unresolved microscopic interactions that may arise in experimental systems, while the overall framework remains a dry active-matter description that does not explicitly account for fluid-mediated interactions \cite{chamolly2026self}.\\
The passive particles are immersed in a bath of chiral active particles that self-propel at a constant speed ($v_0$). To account for natural variations within the active bath, each particle is assigned an intrinsic chirality drawn from a log-normal distribution. By systematically varying the size ratio $S$ and the packing fraction of active particles $\phi_a$, we identify the parameter regimes that support persistent cluster rotation and clarify the roles of cluster geometry, active-bath-induced torque, and chirality heterogeneity in determining the emergence and stability of this collective dynamical state.\\
The rest of the paper is organized as follows: Section~\ref{secII} describes the model and numerical details. Section~\ref{secIII} presents the results. Finally, Section~\ref{secIV} summarizes the main findings and discusses their broader implications.
\section{Model and Numerical Details\label{secII}}
\label{sec:Model}
\begin{figure*}[!hbtp]
\centering
\includegraphics[width=16.5cm, height=6.5cm]{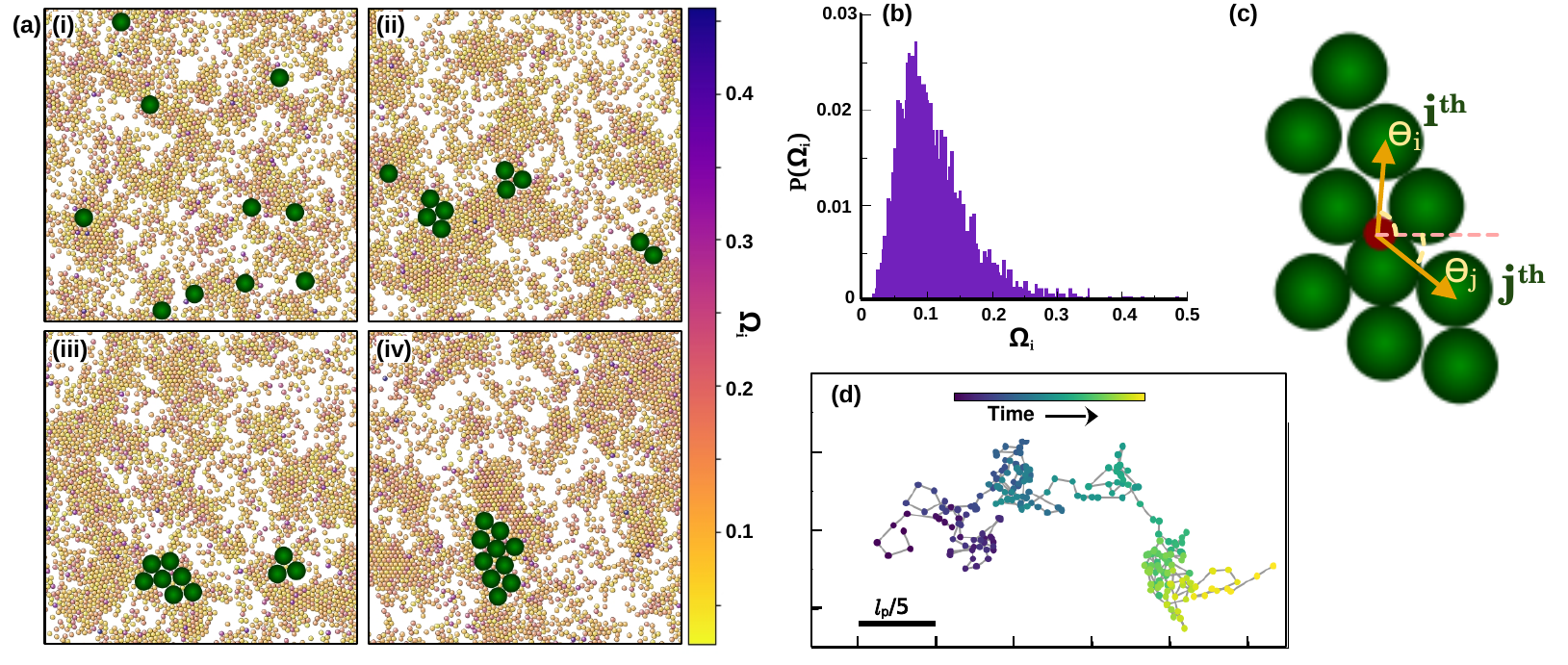}
\caption{(a) Evolution of the system from an initially homogeneous distribution of active and passive particles to a late-time clustered state. Passive particles are shown in green, while chiral active particles are color-coded according to their intrinsic chirality $\Omega_i$. The geometric parameters for the chiral active-passive mixture are $L = 150 a_1$, $a_1 =0.1$, $a_2 =0.4$, $\phi_a = 0.4$, $S=4$. (b) Log-normal probability distribution $P(\Omega_i)$ of chirality for active particles. (c) Snapshot of a passive cluster with the center of mass (COM) highlighted by a red circle. (d) Time trajectory of the cluster's center of mass (COM). (Multimedia available online)}
\label{fig1}
\end{figure*}
We consider a two-dimensional system consisting of a binary mixture of $N_a$ small chiral active particles (cABPs) of radius $a_1$ and $N_p$ large passive particles of radius $a_2$, where $a_2 > a_1$ and $N_a \gg N_p$. The system is confined to a square box of side length $L$ with periodic boundary conditions. Active and passive particles are modeled as disks described by their positions $\mathbf{r}_i^a$ and $\mathbf{r}_i^p$, respectively, the orientation angle of the active particle $\theta_i^a$, along with the self-propulsion direction $\hat{\mathbf{n}}_i = (\cos \theta_i^a, \sin \theta_i^a)$.
The overdamped Langevin equation governs their dynamics.
\begin{equation}
\frac{d\mathbf{r}_i^a}{dt} = v_0 \hat{\mathbf{n}}_i + \mu_1 \sum_{j \neq i} \mathbf{F}_{ij},
\end{equation}
\begin{equation}
\frac{d\theta_i^a}{dt} = -\gamma \sum_{j \in \mathcal{P}} \sin(\theta_i^a - \theta_{ij}) + \sqrt{2D_r} \,\eta_i(t) + \Omega_i,
\end{equation}
where $v_0$ is the self-propulsion speed and $\mu_1$ is the mobility of an active particle. In the orientation update equation (Eq. 2), the sum runs over all passive particles $j$ in contact with the $i^{th}$ active particle. The parameter $\gamma$ determines the strength of the torque induced by these passive particles, and the positional angle $\theta_{ij} = \arctan(\frac{y^a_i - y^p_j}{x^a_i - x^p_j})$ defines their relative orientation. Additionally, $D_r$ is the rotational diffusion constant, and $\eta_i(t)$ is a Gaussian white noise of unit variance that satisfies $\langle \eta_i(t) \eta_j(t') \rangle = \delta_{ij} \delta(t - t')$. Here, $\delta_{ij}$ and $\delta(t - t')$ indicate that the noise is uncorrelated between different particles and at different times, respectively.\\
While the noise terms capture random fluctuations, the intrinsic rotational dynamics of the active particles must also be carefully considered. Experimental studies of microswimmers, such as {\it E. coli}, reveal that physical dimensions and kinematic properties (such as cell length, swimming speed, and rotational dynamics) exhibit broad, naturally right-skewed distributions due to inherent population inhomogeneities \cite{chattopadhyay2006swimming,wilson2011differential,gangan2017threshold,lisicki2019swimming}. To mimic this natural inhomogeneity in our model's rotational dynamics, we assume that the intrinsic chirality $\Omega_i$ of the particles is drawn from a log-normal distribution with a mean $\Omega_0$ and  logarithmic standard deviation $\sigma$,
 $P(\Omega_i) =\dfrac{1}{\Omega_i\sigma\sqrt{2\pi}} \text{exp}\Big(-\dfrac{(\ln \Omega_i-\ln\Omega_0)^2}{2\sigma^2}\Big)$. We additionally checked the results against a baseline model where a constant chirality was maintained for all cABPs (Sec.~\ref{subsec:cd}).\\
The position $\mathbf{r}_i^p$ of the passive particles evolves according to:
\begin{equation}
\frac{d\mathbf{r}_i^p}{dt} = \mu_2 \sum_{j \neq i} \mathbf{F}_{ij}^{pp} + \mu_2 \sum_{k} \mathbf{F}_{ik},
\end{equation}
where $\mu_2$ is the mobility of a passive particle. The terms $\mathbf{F}_{ij}^{pp}$ and $\mathbf{F}_{ik}$ correspond to the passive-passive and active-passive interaction forces, respectively. To model volume exclusion, both the active-active and active-passive interactions share the same mathematical form. Thus, the general soft repulsive interaction force $\mathbf{F}_{ij} = F_{ij} \hat{\mathbf{r}}_{ij}$ exerted on any particle $i$ by any particle $j$ is defined as:
\begin{equation}
{F}_{ij}=
\begin{cases}
k ((a_{i} + a_{j})-{r}_{ij}),  & {r}_{ij} \leq (a_{i} + a_{j}) \\
0, & \text{otherwise},
\end{cases}
\end{equation}
where $r_{ij}=|\mathbf{r}_i-\mathbf{r}_j|$, $a_i$, and $a_j$ assume the value $a_1$ for active particles and $a_2$ for passive particles. Here, $k$ is the repulsion stiffness parameter used to maintain the effective volume-exclusion interaction. The passive-passive interaction is given by:
\begin{equation}
F_{ij}^{pp} =
\begin{cases}
k \left(2a_{2}-{r}_{ij}\right),
& {r}_{ij}\ \leq 2a_{2} \\
\epsilon_{pp} \left(2a_{2}-{r}_{ij}\right), &2a_{2} < {r}_{ij}\leq 2a_{2} + S a_1 \\
0, & {r}_{ij} > 2a_{2} + S a_1,
\end{cases}
\end{equation}
where $2a_2$ represents the sum of the radii of two interacting passive particles. The parameter $S = \frac{a_2}{a_1}$ is the size ratio and is varied in the range $[1,6]$. The variation of force is shown in the Appendix~\ref{sec:force}.\\
To determine the optimal clustering behavior, we explicitly tested varying passive-passive attraction strengths, including a weaker attraction of $\epsilon_{pp}= 0.1k$ and a stronger attraction of $\epsilon_{pp}=0.6k$ (see Appendix~\ref{sec:force}). We selected an attraction strength of $\epsilon_{pp}=0.3k$ for our primary investigations to strike a crucial dynamic balance. As observed in our simulations, a weak attraction ($\epsilon_{pp}=0.1k$) is insufficient to stabilize the passive clusters against the strong collisions and fluctuations of the active bath. Conversely, a strong attraction ($\epsilon_{pp}=0.6k$) leads to rigid, kinetically arrested structures that inhibit internal particle rearrangements and neighbor exchanges.\\
In recent experiments \cite{semwal2022tunable}, the effective depletion-induced attraction between colloids is tuned by varying the polymer concentration in the mixture. It has been found that high depletion attractions cause colloidal suspensions to arrest into rigid gel states, while intermediate attraction strengths allow for the formation of stable clusters that avoid irreversible gelation.
\begin{figure*}[!hbtp]
\centering
\includegraphics[width=17cm, height=12cm]{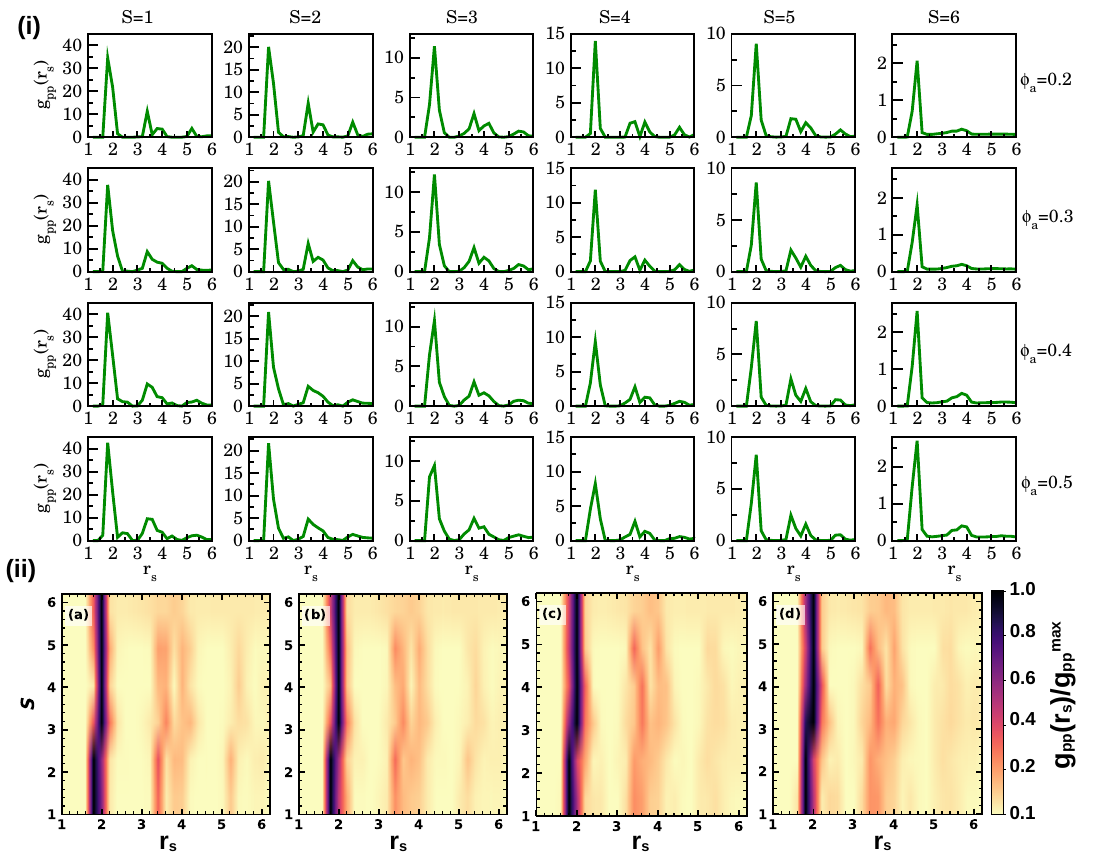}
\caption{(i) Passive-passive radial distribution function $g_{pp}(r_s)$ as a function of the scaled separation $r_s=r^p/(Sa_2)$. Columns correspond to the size ratios $S=1,2,3,4,5,$ and $6$, while rows correspond to the active-particle area fractions $\phi_a=0.2,0.3,0.4,$ and $0.5$. (ii) Heat-map representation of the normalized passive-passive radial distribution function, $g_{pp}(r_s)/g_{pp}^{\max}$, as a function of $r_s$ and $S$ for (a) $\phi_a=0.2$, (b) $\phi_a=0.3$, (c) $\phi_a=0.4$, and (d) $\phi_a=0.5$.(Multimedia available online)}
\label{fig2}
\end{figure*}
Simulations begin with a random distribution of active and passive particles in a square box of $150a_1 \times 150a_1$, with random initial velocity directions assigned to the active particles, ensuring non-overlapping initial configurations. The area fraction of active particles, $\phi_a = \frac{N_a \pi a_1^2}{L^2}$, is varied in the range $[0.2,0.5]$, and the number of passive particles is fixed at $10$. Other parameters are kept fixed at   $v_0=0.50$, $D_r=0.1$, $\gamma = 1.0$, and $\Omega_i$,  following a $\log$-normal distribution with a mean $\Omega_0 =0.11$, as shown in (Fig.~\ref{fig1} (b)) (Multimedia available online). The persistence length is $\ell_p = v_0/D_r$ and the persistence time is $\tau_p = a_1/v_0$. The small integration time step is $\Delta t = 5\times 10^{-4}\tau_p$. All lengths and time scales are measured in units of cABP size $a_1$ and persistence time $\tau_p$, respectively. The above equations describe simulations run for $T=10^4\tau_p$ time steps. A single simulation step is counted once all active particles' positions and orientations are updated. Observations are performed after $5\times 10^3\tau_p$, when the steady state is reached. The steady state is characterized by the absence of statistical patterns in the particle's dynamics. The averaging is performed over a total of $5\times 10^3\tau_p$ times in the steady state and over $200$ to $300$ independent initial realizations.\\ 
This model captures the interplay between the active particle packing fraction and the size ratio while incorporating fixed parameters for volume exclusion, activity, torque induced by passive particles, and the weak passive-passive attraction. As observed in recent experiments on active-passive mixtures, the size ratio between passive and active particles, along with the density of the active particles, is a key parameter \cite{jiang2025experimental,kushwaha2023phase}. Based on these observations, we systematically vary the size ratio $S$ and the active packing fraction $\phi_a$ to investigate their effects in mixtures of cABPs and passive particles. Additionally, we vary the standard deviation of the chirality distribution for the cABPs using $\sigma=0$ (constant chirality), $\sigma=0.2$, and $\sigma=0.47$. Most of the results are obtained for $\sigma=0.47$. 
\section{Results \label{secIII}}
We begin by showing the time evolution of the chiral active-passive mixture from an initially homogeneous state to the clustered dynamical state that forms at late times. Fig.~\ref{fig1} (a) (Multimedia available online) shows representative time-ordered snapshots of the system, where the active and passive particles are initially distributed randomly and remain spatially dispersed but progressively reorganize as the dynamics evolve. As time increases, the passive particles (shown in green) aggregate due to the effective interactions generated by the surrounding active bath, eventually forming a compact cluster embedded in a sea of chiral active particles. These surrounding active particles are color-coded according to their intrinsic chirality $\Omega$, which follows a log-normal distribution with a logarithmic standard deviation $\sigma=0.47$, as shown in Fig.~\ref{fig1} (b) (Multimedia available online).\\
This late-time clustered state is not static; rather, it shows persistent collective translational and rotational dynamics.\\
To characterize this motion quantitatively, we introduce the geometrical observables used throughout the rest of the paper. The center of mass (COM) \cite{bai2008calculating} of the passive cluster is defined as the average position of all passive particles $\mathbf{r}_{CM}=\sum_i^{N_p}\mathbf{r}^{p}_i/N_p$, where $N_p$ is the number of passive particles in the cluster. The angular position of each passive particle is then measured with respect to this COM as $\theta_i(t)=tan^{-1}\big(\dfrac{y^p_i-y_{CM}}{x^p_i-x_{CM}}\big)$, which allows us to track the rotational dynamics of the cluster. The cluster geometry and the COM construction are illustrated in Fig.~\ref{fig1} (c) (Multimedia available online), while the laboratory-frame trajectory of the cluster COM shown in Fig.~\ref{fig1} (d) (Multimedia available online) exhibits a persistently curved path, indicating that the passive cluster undergoes simultaneous translation and rotation over a long period.\\
We first discuss the characteristics of the passive cluster by examining its internal structural properties through the passive-passive radial distribution function \cite{dolai2018phase} ($g_{pp}(r_s)$) as a function of the scaled distance ($r_s=r^p/(Sa_2)$), as shown in Fig.~\ref{fig2} (Multimedia available online). For intermediate size ratios ($S=3,4,$ and $5$) across all investigated packing fractions ($\phi_a=0.2,0.3,0.4,$ and $0.5$), $g_{pp}(r_s)$ exhibits higher-order peaks at $r_s\approx2\sqrt{3},4$, characteristic of hexagonal close-packed order. This indicates that the steady-state cluster retains local hexagonal ordering in the rotating regime, consistent with the dynamics observed in Movies $3$, $4$, and $5$, and with the real-space snapshots shown in Appendix~\ref{sec:app}.
The corresponding heat-map representation in Fig.~\ref{fig2}(ii) summarizes these trends across the full parameter range. The dominant near-contact peak remains centered at $r_s\approx2$, whereas the higher-order coordination shells are well defined for intermediate size ratios, weaken for $S\le2$, and disappear almost completely at $S=6$, consistent with the loss of local hexagonal order.\\
For small size ratios ($S=1,2$) at high packing fractions ($\phi_a=0.4,0.5$), the passive and active particles are comparable in size, allowing active particles to intermittently penetrate or become trapped within the passive aggregate, thereby disrupting the local packing (see Appendix~\ref{sec:app}). At $S=6$, the passive aggregate frequently forms and breaks apart. We therefore focus our analysis of the cluster dynamics on size ratios up to $S=5$, where the aggregates remain stable over a finite timescale.\\
To understand cluster dynamics in the active bath, we first characterize the structural morphology of the passive clusters and their mechanical response to the surrounding medium.\\
We quantify morphological changes in cluster shape using the Gyration tensor \cite{rubinstein2003polymer}, $\mathcal{R}(t)$, defined as $\mathcal{R}(t)=\dfrac{1}{N_p}\sum_{i}^{N_p} \mathbf{r}^{p}_{i, CM}(t) \otimes \mathbf{r}^{p}_{i, CM}(t)$, where $\mathbf{r}^{p}_{i, CM}(t)= \mathbf{r}^
{p}_{i}(t)-\mathbf{r}_{CM}(t)$ and $\otimes$ denotes the tensor product \cite{rubinstein2003polymer}. In matrix form,
\begin{equation*}
\mathcal{R}(t)= 
\begin{bmatrix}
R_{xx}(t) & R_{xy}(t)\\
R_{yx}(t) & R_{yy}(t)
\end{bmatrix}  
\end{equation*}
The components of the gyration tensor are given by;
\begin{align*}
R_{xx}(t) &= \frac{1}{N_p}\sum_{i=1}^{N_p} (x_{i,CM}^{p}(t))^{2},\\
R_{yy}(t) &= \frac{1}{N_p}\sum_{i=1}^{N_p} (y_{i,CM}^{p}(t))^{2},\\
R_{xy}(t) &= R_{yx}(t)=\frac{1}{N_p}\sum_{i=1}^{N_p} x_{i,CM}^{p}(t)y_{i,CM}^{p}(t).
\end{align*}
Here, $x_{i,CM}^{p}(t)$ and $y_{i,CM}^{p}(t)$ are the $x$ and $y$ components of the position vector (${\bf r}_i^{p}$) of the $i^{th}$ passive particle relative to the COM. The squared radius of gyration at time $t$ is $R_g^2(t)\equiv Tr(\mathcal{R}(t))=\lambda_1(t)+\lambda_2(t)$, where $\lambda_1$ and $\lambda_2$ are the eigenvalues of the gyration tensor. Another shape measure is the asphericity $A_p$ \cite{paoluzzi2016shape,aronovitz1986universal}, defined as; 
\begin{equation*}
A_p=\big<\dfrac{(\lambda_1(t)-\lambda_2(t))^2}{(\lambda_1(t)+\lambda_2(t))^2}\big>    
\end{equation*}
where $<...>$ denotes the time-averaged data during the steady state and across $200$ independent realizations. $A_p$ varies between $0$ and $1$, which correspond to a circle and a rod, respectively. Thus, asphericity quantifies the deviation of a cluster’s shape from a perfectly circular or isotropic configuration, with higher values indicating more elongated or anisotropic configurations.
\begin{figure}[H]
\centering
\includegraphics[width=8.5cm, height=7.5cm]{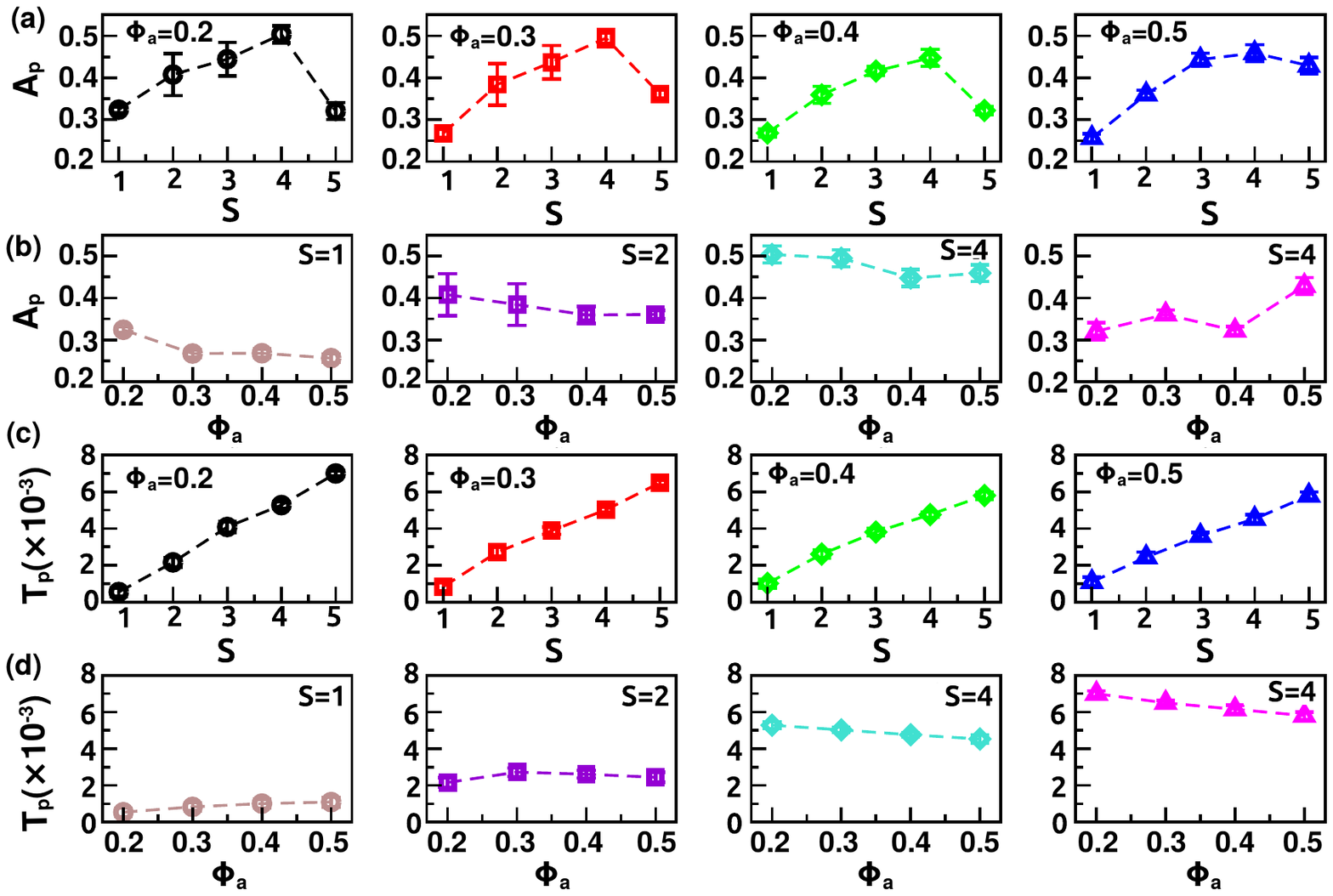}
\caption{(a) Mean cluster asphericity ($A_p$) as a function of the size ratio ($S$) for different active packing fractions ($\phi_a$). (b) Mean cluster asphericity ($A_p$) as a function of ($\phi_a$) for different size ratios ($S$). (c) Mean magnitude of the net torque ($T_p$) acting on the passive cluster as a function of ($S$) for different ($\phi_a$). (d) Mean magnitude of the net torque ($T_p$) acting on the passive cluster as a function of ($\phi_a$) for different ($S$), as indicated in the legend. All quantities are averaged over the clustered steady state and over independent realizations. (Multimedia available online; see Movie $1$ and $2$)}
\label{fig3}
\end{figure}
Fig.~\ref{fig3} (a) (Multimedia available online) presents the variation of cluster asphericity ($A_p$) as a function of the size ratio ($S$) for different packing fractions ($\phi_a$). For each size ratio, asphericity is averaged over time during the cluster's persistent rotational phase. The plot reveals a non-monotonic trend at intermediate size ratios $(S=3,4)$ across all $\phi_a$. At small size ratios ($S=1,2$), asphericity remains low, indicating that the clusters are nearly circular and isotropic. As the size ratio increases to intermediate values ($S=3,4$), asphericity increases substantially, indicating a more elongated or irregular cluster shape. For larger clusters ($S=5$), the asphericity decreases again, indicating a return to more symmetric shapes; the same is seen in Fig.~\ref{fig8} (Multimedia available online). Although asphericity shows a strong dependence on the size ratio ($S$), it exhibits a weak dependence on the packing fraction ($\phi_a$), as shown in Fig.~\ref{fig3} (b) (Multimedia available online). This behavior demonstrates that intermediate-sized clusters adopt more elongated, anisotropic average shapes. In contrast, small($S=1,2$) or large clusters($S=5$) maintain more isotropic, symmetric configurations, and this structural trend remains largely insensitive to the packing fraction of active particles.\\
\begin{figure*}[!hbtp]
\centering
\includegraphics[width=17cm,height=12cm]{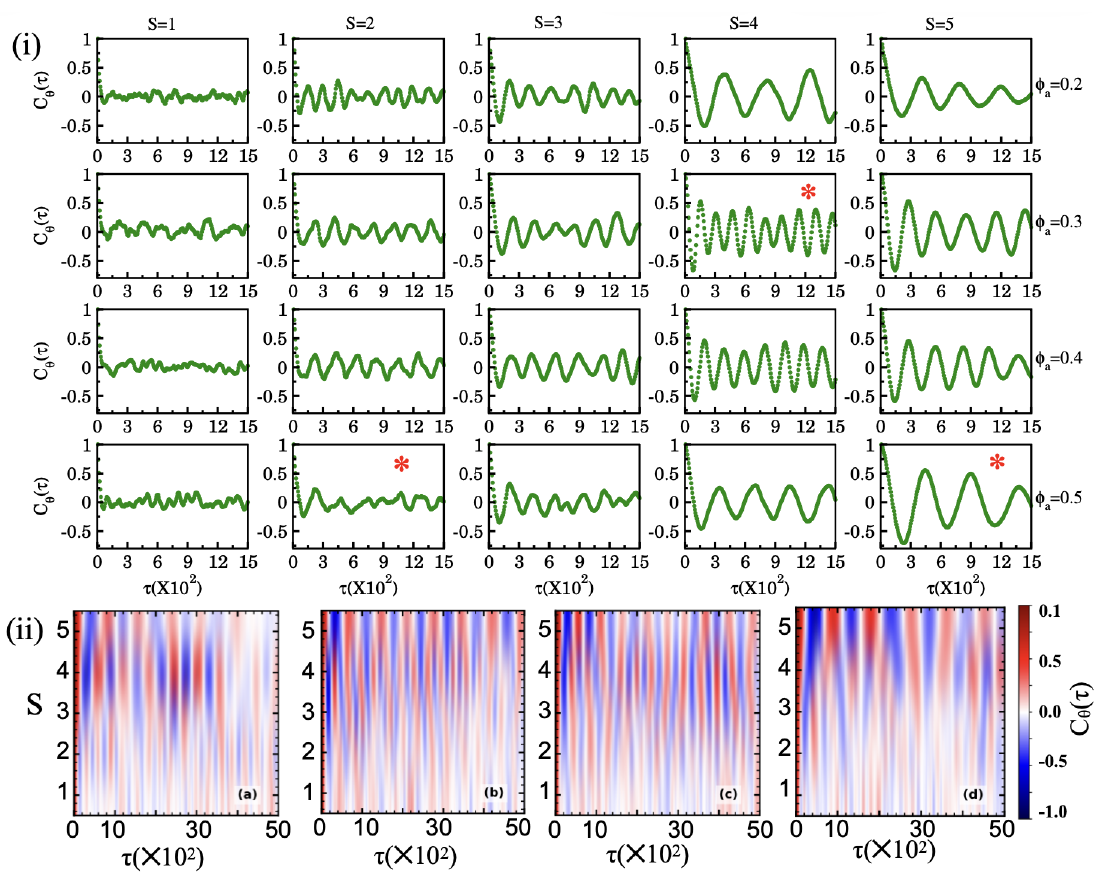}
\caption{(i) Angular autocorrelation function $C_\theta(\tau)$ as a function of the lag time $\tau$. Rows correspond to the active-particle area fractions $\phi_a$, while columns correspond to the size ratios $S$, as labeled in the panels. The $x$-axis denotes time in units of $\tau_p$. The Movies corresponding to the parameter sets marked by stars (\textcolor{red}{*}) are provided in Appendix~\ref{sec:anim}. (ii) Heat-map representation of the angular autocorrelation function $C_\theta(\tau)$ as a function of the lag time $\tau$ and size ratio $S$ for (a) $\phi_a=0.2$, (b) $\phi_a=0.3$, (c) $\phi_a=0.4$, and (d) $\phi_a=0.5$. Red and blue regions denote positive and negative angular correlations, respectively. (Multimedia available online; see Movies $3$-$5$.)}
\label{fig4}
\end{figure*}
To examine whether these shape fluctuations correlate with mechanical forcing from the active bath, we compute the magnitude of the net torque $(T_p)$ \cite{torrik2021dimeric} on the passive cluster, $T_p \equiv \left<\tilde{T}_p\right>, \quad\tilde{T}_p=|\sum_{i=1}^{N_p} \left(\mathbf{r}_i - \mathbf{r}_{CM}\right) \times \mathbf{F}_i^{p}|$,
where $<...>$ denotes the time-averaged data during the steady state and across $200$ independent realizations, and $\mathbf{F}_i^{p}$ is the total force on the $i^{\mathrm{th}}$ passive particle due to all neighboring active and passive particles. \\
Panels (c,d) of Fig.~\ref{fig3} (Multimedia available online) show that the averaged net torque magnitude $T_p$ on the passive cluster increases monotonically with the size ratio $S$ across all active packing fractions $\phi_a$. Up to $S=4$, this increase in net torque parallels the enhanced cluster asphericity $A_p$, as more elongated shapes experience asymmetric forcing from the surrounding chiral active particles. However, at $S=5$, the decrease in $A_p$ contrasts with the continued increase in $T_p$. The large size of the passive cluster alters the mechanical response. Instead of driving coherent cluster rotation, the large torque at $S = 5$ induces internal particle rearrangements within the cluster, where passive particles exchange neighbors and shuffle positions. This is evident from the changes in local coordination at $S = 5$ compared to the stable contacts seen at $S = 4$, indicating that the mechanical input is dissipated into positional dynamics rather than cluster rotation (see Appendix~\ref{sec:nbd}).\\
To understand how these structural properties and mechanical forces relate to the cluster's angular dynamics, we compute the angular autocorrelation function of passive particles relative to the cluster COM,
$C_\theta(\tau)=\langle\cos[\delta\theta_i(t)-\delta\theta_i(t+\tau)]\rangle$, where $\langle\cdots\rangle$ denotes the average over all passive particles. The fluctuations in $\theta_i$ are defined as $\delta\theta_i(t)=\theta_i(t)-\bar{\theta_i}$, where $\bar{\theta_i}$ is the mean value of $\theta_i(t)$ over time. The results shown in Fig.~\ref{fig4} (Multimedia available online) systematically map this function across varying active packing fractions ($\phi_a$) and size ratios ($S$).\\
For small size ratios ($S=1,2$) across all $\phi_a$, the passive clusters maintain nearly circular, isotropic configurations with low asphericity. Coinciding with this structural symmetry, the measured net torque is relatively weak, and $C_\theta(\tau)$ exhibits a slow, weakly oscillatory, or noisy decay (especially at $S=1$, where particle sizes are comparable). This lack of collective rotation is visualized in Appendix~\ref{sec:anim} (Movie $3$), which corresponds to the starred panel at $S=2$ and $\phi_a=0.5$. As the size ratio increases to intermediate values ($S=3,4$), the enhanced geometric anisotropy corresponds to an increased asymmetric torque (as discussed in Fig.~\ref{fig3} (Multimedia available online)). However, the rotational signatures differ between the two sizes. For $S=3$, the oscillations in $C_\theta(\tau)$ are present but less pronounced, consistent with its high translational diffusivity (see Fig.~\ref{fig6} (Multimedia available online)), because the cluster undergoes simultaneous translation and rotation, making the angular signal noisier. In contrast, for $S=4$, $C_\theta(\tau)$ displays clear periodic oscillations with deep negative dips, indicating that the cluster completes full revolutions before its orientational memory decorrelates. Appendix~\ref{sec:anim} (Movie $4$), corresponding to the starred panel at $S=4$ and $\phi_a=0.3$, illustrates this behavior. For the largest size ratio ($S=5$), the cluster exhibits slower oscillations that persist over longer time scales. Although the measured net mechanical torque reaches its highest magnitude, concurrent internal particle rearrangements and neighbor exchanges increase the time required to complete a rotation. The corresponding heat-map representation in Fig.~\ref{fig4}(ii) summarizes these trends across the investigated parameter space. Alternating positive and negative correlation bands are most clearly observed for the intermediate size ratios ($S\approx3-5$), whereas they become progressively weaker for smaller size ratios, consistent with the reduced rotational coherence inferred from the autocorrelation curves.\\
While the rotational behavior depends primarily on the size ratio, the active packing fraction $\phi_a$ modulates the active driving forces and the mobility of the cluster. At the lowest density ($\phi_a=0.2$), the measured torque is comparatively weak, leading to slower decorrelation in $C_\theta(\tau)$ and smaller differences among the various cluster sizes. Increasing the packing fraction to $\phi_a=0.3$ and $0.4$ enhances the active torque, making the oscillatory behavior more evident and increasing the rotation rate. At the highest packing fraction ($\phi_a=0.5$), the crowded environment restricts the cluster mobility and reduces the measured torque relative to its maximum at intermediate densities. Consequently, the oscillations become slower, and their amplitude decreases compared to $\phi_a=0.4$. These results identify intermediate size ratios ($S=3,4$) together with moderate active packing fractions ($\phi_a=0.3,0.4$) as the parameter regime most favorable for sustained collective rotation. These observations are further corroborated by the heat maps in Fig.~\ref{fig4}(ii), which show the strongest oscillatory correlation bands for intermediate size ratios at moderate packing fractions ($\phi_a=0.3,0.4$), while weaker bands at low and high densities reflect the reduced rotational coherence.
\begin{figure*}[!hbtp]
\centering
\includegraphics[width=15cm, height=10cm]{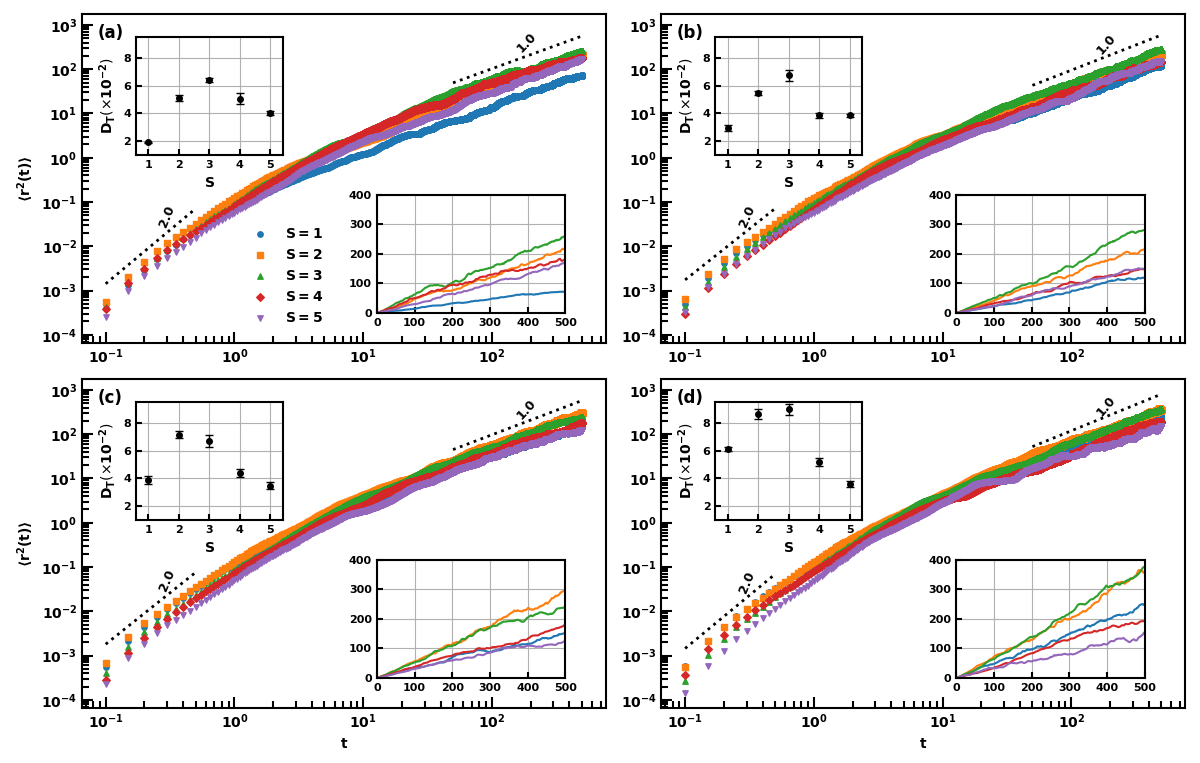}
\caption{Mean squared displacement $\langle \Delta \mathbf{r}^2(t) \rangle$ vs. $t$ (in units of $\tau_p$) of the passive cluster for different values of $S$, as indicated in the legend. Each panel corresponds to a fixed value of the packing fraction of active particles $\phi_a$, as indicated in the panel labels: (a) $\phi_a=0.2$, (b) $\phi_a=0.3$, (c) $\phi_a=0.4$, and (d) $\phi_a=0.5$. Insets show the same data on linear scales over the same time intervals, as well as the effective late-time diffusivity $D_T$ vs. $S$. (Multimedia available online)}
\label{fig6}
\end{figure*}
\begin{figure*}[!hbtp]
\centering
\includegraphics[width=15cm, height=10cm]{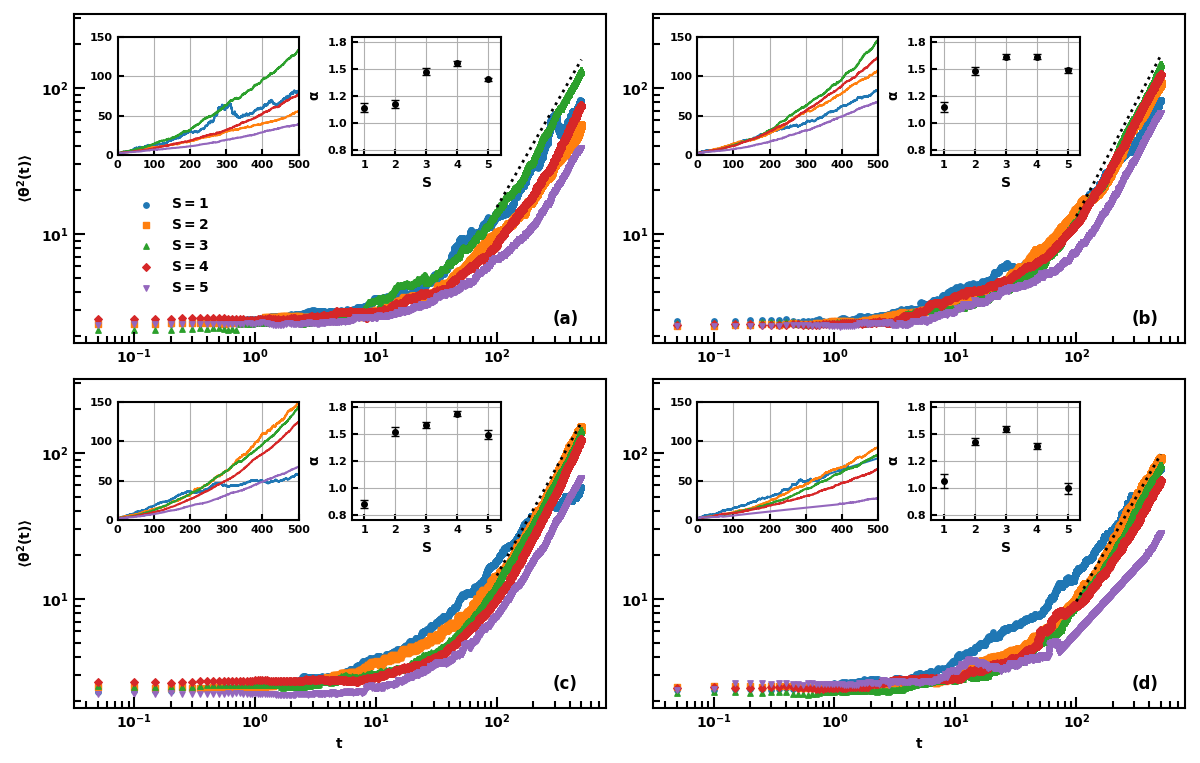}
\caption{Mean squared angular displacement $\langle \Delta \theta^2(t) \rangle$ of the passive cluster plotted as a function of time $t$ (in units of $\tau_p$) for different size ratios $S$. The dashed lines indicate a slope of $1.5$. Each panel corresponds to a fixed $\phi_a$, as (a) $\phi_a=0.2$, (b) $\phi_a=0.3$, (c) $\phi_a=0.4$, and (d) $\phi_a=0.5$. Insets show the same data on linear scales over short time intervals, as well as the exponent $\alpha$ vs. $S$. (Multimedia available online)}
\label{fig5}
\end{figure*}
While the autocorrelation reveals the temporal memory of angular motion, it does not, by itself, distinguish among subdiffusive, diffusive, and superdiffusive rotation. To quantify the nature of the rotational dynamics more precisely, we compute the mean-squared angular displacement (MSAD) of the cluster, Fig.~\ref{fig5} (Multimedia available online) $\langle\Delta\theta^2(t)\rangle=\langle\frac{1}{N_{p}} \sum_{i=1}^{N_{p}}[\Delta\theta_i(t,t_0)]^2\rangle$, where $\Delta\theta_i(t, t_0) = \theta_i(t+t_0) - \theta_i(t_0)$ represents the continuous, unwrapped angular displacement of the $i^{th}$ passive particle relative to the cluster's COM over a time lag $t$, and $N_p$ is the total number of passive particles. Next, we evaluate the translational motion of the passive cluster (Fig.~\ref{fig6} (Multimedia available online)) by calculating the translational mean-squared displacement (MSD) of its COM, $\langle \Delta \mathbf{r}^2(t) \rangle=\langle[\mathbf{r}_{CM}(t+t_0)-\mathbf{r}_{CM}(t_0)]^2\rangle$, where $\mathbf{r}_{CM}(t)$ is the unwrapped spatial position vector of the cluster's COM at time $t$ \cite{jhajhria2025kinetics}. In both the MSAD and MSD equations, $\langle\dots\rangle$ denotes the average over all 200 independent realizations and many reference times $t_0$.\\
Fig.~\ref{fig5} (a-d) (Multimedia available online) shows the MSAD ($\langle\Delta\theta^2(t)\rangle$) of the passive cluster for increasing $\phi_a$. In each panel, the main plot presents the MSAD on log-log scales. At the same time, the insets display the same data on linear axes, and the corresponding late-time exponent $\alpha$, extracted from the power-law relation $\langle\Delta\theta^2(t)\rangle \sim t^\alpha$, is plotted as a function of the size ratio $S$. Across all packing fractions, the MSAD exhibits a clear crossover from an early-time flat trend to a late-time faster variation. Extracting this exponent $\alpha$ provides a quantitative measure of the rotational dynamics.\\ 
 Within the optimal window of $S=3$ and $4$, the system exhibits superdiffusive angular motion, confirming the emergence of sustained, coherent rotation. Consistent with the trends discussed for the autocorrelation $C_\theta(\tau)$, the exponent $\alpha$ also depends on the packing fractions. Specifically, $\alpha$ reaches its maximum value ($\alpha \simeq 1.7$) at intermediate packing fractions ($\phi_a = 0.3$ and $0.4$), while it is notably lower for both smaller ($\phi_a = 0.2$) and larger ($\phi_a = 0.5$) packing fractions.\\
Conversely, outside the optimal window, the angular dynamics are generally weaker. For $S=1$ across all $\phi_a$, the MSAD is notably noisy (inset of Fig.~\ref{fig5} (Multimedia available online)). Because the passive particles are comparable in size to the active particles, the cluster fails to establish a stable rotational axis, which is reflected in the rapid decorrelation of $C_\theta(\tau)$. For $S=2$ and $5$, the MSAD curves are smoothly resolved, and the rotational dynamics follow a similar $\phi_a$ dependence as in $S=3$ and $4$. The motion approaches superdiffusive at intermediate packing fractions ($\phi_a = 0.3$ and $0.4$) but remains close to diffusive at extreme packing fractions ($\phi_a = 0.2$ and $0.5$). However, the underlying physical reasons for their sub-optimal rotation differ. For $S=2$, the clusters assemble into highly isotropic and nearly circular shapes, leading to weaker net torques. In contrast, at the largest size ratio ($S=5$), the intense active forces induce internal particle rearrangements rather than clean rigid-body rotation, slightly suppressing the maximum superdiffusive regime.\\
Having established the conditions for sustained collective rotation, we now examine how the surrounding active bath drives the translational motion of the passive cluster. Fig.~\ref{fig6} (a–d) (Multimedia available online) shows the MSD of the cluster COM, $\langle \Delta r^2(t) \rangle$, for increasing $\phi_a$. Across all $\phi_a$, the MSD exhibits a clear crossover from an initial ballistic-like regime to a linear long-time regime, indicating purely diffusive translational transport at late times. This transition reflects transport signatures characteristic of active Brownian particles \cite{howse2007self}. The insets display the corresponding effective long-time diffusivity, $D_T = \lim_{t \to \infty} \langle \Delta r^2(t) \rangle / 4t$, as a function of the size ratio $S$.\\
A consistent trend emerges across all packing fractions $\phi_a$. The translational diffusivity $D_T$ is maximized for intermediate size ratios $S=2$ and $3$, but it noticeably decreases for both the small ($S=1$) and the larger  ($S=4$ and $5$). This behavior can be directly linked to the clusters' geometric morphology and their structural response to the active bath. For the smallest size ratio ($S=1$), the active and passive particles are comparable in size. As seen in Fig.~\ref{fig8} (Multimedia available online), active particles frequently penetrate and disrupt the passive aggregate rather than push against its surface. This continuous structural disturbance corresponds to a lack of cohesion, leading to inefficient COM transport and a lower $D_T$. As $S$ increases to $2$ and $3$, the clusters form more stable configurations. While $S=2$ is highly isotropic, $A_p$ begins to increase at $S=3$ [Fig.~\ref{fig4} (a) (Multimedia available online)]. However, the net torque $T_p$ in this regime has not yet reached the high values seen in larger clusters. Because the asymmetric forcing is not yet strong enough to channel purely into rigid-body rotation, the random collisions from the active bath contribute significantly to COM fluctuations, driving the highest observed values of $D_T$.\\
For $S = 4$, the highly anisotropic cluster geometry corresponds to a peak in rotational dynamics (Fig.~\ref{fig5} (Multimedia available online)) and a significant decrease in translational diffusivity $D_T$. This indicates that the active bath's forcing predominantly drives collective rotational motion rather than center of mass (COM) translation.\\
For the largest size ratio $S=5$, $D_T$ remains low. At this scale, the aggregate's large size, combined with intense active forcing, leads to enhanced internal particle rearrangements and neighbor exchanges (Appendix~\ref{sec:nbd}). Rather than driving efficient COM transport or rigid-body rotation, the active pushing corresponds to internal positional shifts, keeping both $D_T$ and rotational coherence low.\\
Furthermore, the magnitude of $D_T$ is sensitive to $\phi_a$. As the active bath becomes denser (from $\phi_a = 0.2$ to $0.5$), $D_T$ systematically increases for all $S$. The increased number of active particles colliding with the passive aggregate provides a stronger collective driving force, thereby directly increasing COM fluctuations and enhancing translational transport across all cluster sizes  \cite{mino2011enhanced,wang2020inhibition}. \\
Comparing these translational and angular dynamics reveals a clear separation in both time and parameter space. Temporally, translation dominates the initial dynamics, as evidenced by a ballistic-like rise in the MSD alongside a trapped, slow-growing MSAD (Fig.~\ref{fig5} (Multimedia available online)). For late times, this behavior inverts. For $S=3$ and $4$, the MSAD becomes strongly superdiffusive while the MSD transitions to standard diffusion, showing that collective rotation dominates the long-time behavior. Furthermore, the conditions that maximize these two motions are distinct. Translational diffusion peaks at slightly smaller isotropic size ratios ($S=2, 3$), whereas optimal rotational dynamics require the highly anisotropic geometries of $S=3, 4$.
\subsection{Role of system size and packing fraction of the  passive particles}
\begin{figure}[H]
\centering
\includegraphics[width=6cm, height=5cm]{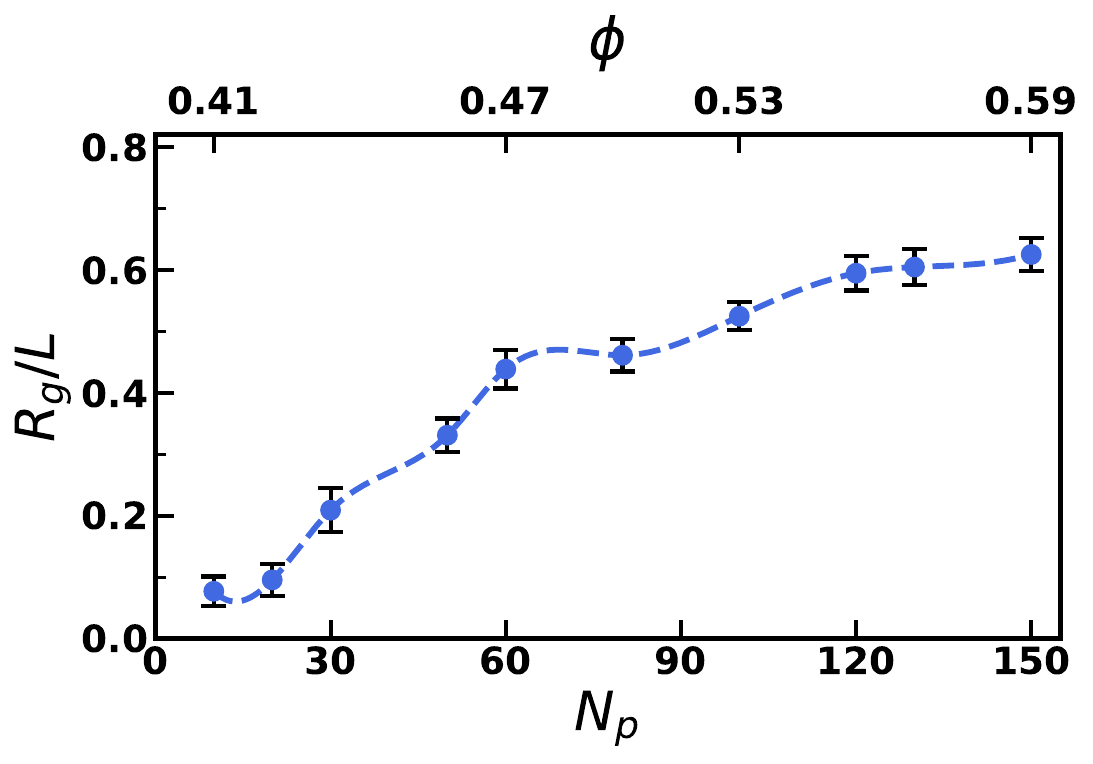}
\caption{Radius of gyration normalized by the system size $R_g/L$ of the passive cluster as a function of the number of passive particles $N_p$ for a system of size $L=200a_1$, $\phi_a=0.4$ and $S=4$. The corresponding total packing fraction $\phi=\phi_a+\phi_p$ is indicated on the upper horizontal axis. Error bars denote the standard deviation obtained from independent realizations. (Multimedia available online)}
\label{fig7}
\end{figure}
\begin{figure}[H]
\centering
\includegraphics[width=7cm, height=10cm]{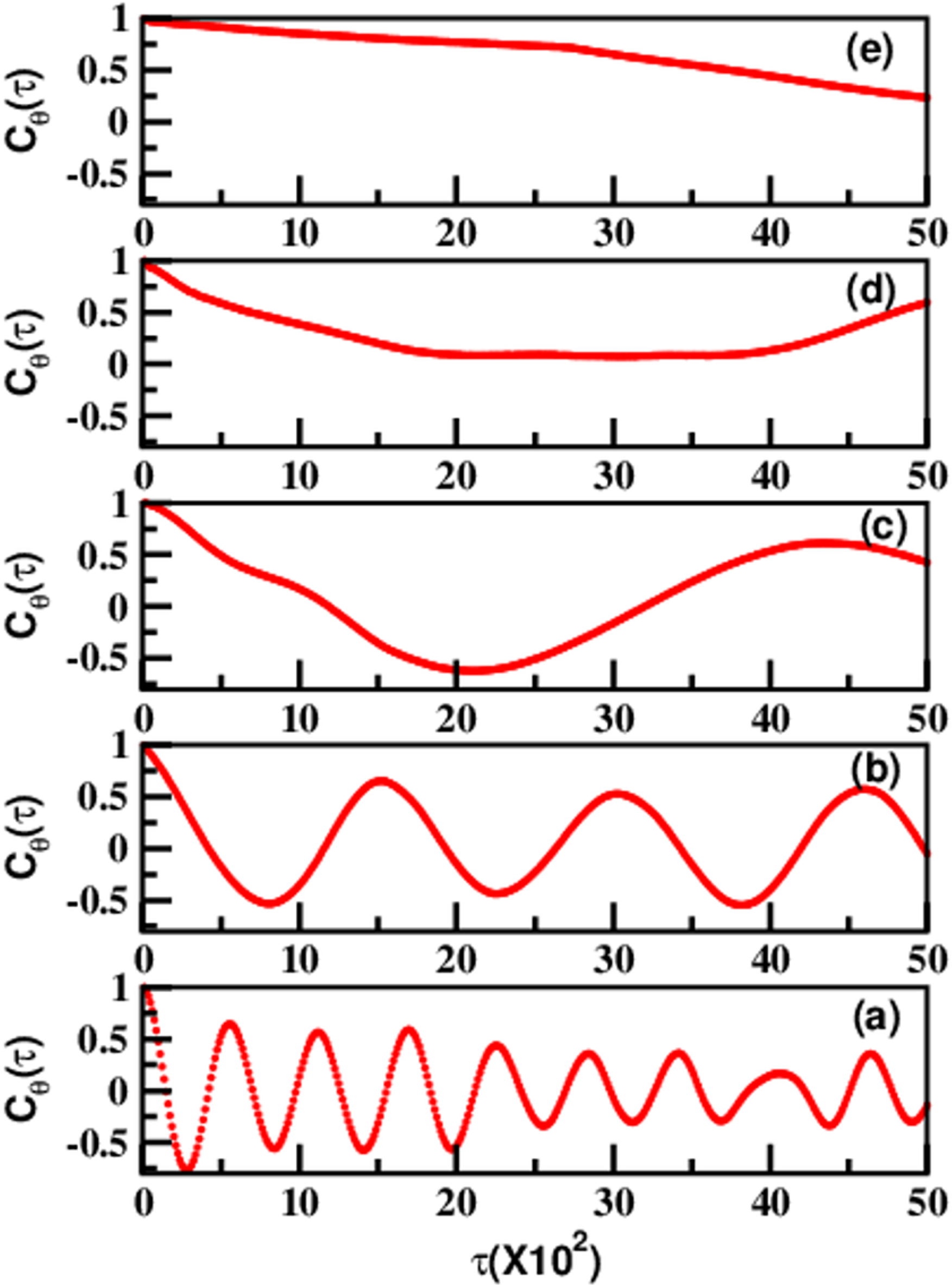}
\caption{Angular autocorrelation function $C_\theta(\tau)$ of the passive cluster for increasing numbers of passive particles in a system of size $L=200a_1$: (a) $N_p=20$, (b) $N_p=30$, (c) $N_p=50$, (d) $N_p=100$, and (e) $N_p=150$.(Multimedia available online; see Movie 12 and 13)}
\label{fig8}
\end{figure}
To examine the role of system size, we performed additional simulations in a larger domain of linear size $L=200a_1$ while increasing the number of passive particles $N_p$. Fig.~\ref{fig7} (Multimedia available online) shows that the normalized radius of gyration $R_g/L$ increases monotonically with $N_p$, indicating the progressive growth of the passive aggregate. The corresponding angular autocorrelation functions are shown in Fig.~\ref{fig8} (Multimedia available online), and representative movies for $N_p=30$ and $N_p=100$ are provided in Appendix~\ref{sec:anim} as Movies $12$ and $13$, respectively. Movie $11$ shows the case $N_p=10$ in the larger domain ($L=200a_1$), confirming that the rotational behavior persists for the original cluster size as well.\\
For the smallest cluster shown here, $N_p=20$, the autocorrelation exhibits clear oscillations, indicating persistent collective rotation. As $N_p$ increases, the aggregate becomes more extended and internally heterogeneous, so the active bath can no longer impose a uniform torque across the full cluster. Local rearrangements then become more important, and the oscillatory signal in $C_\theta(\tau)$ gradually weakens. By $N_p=100$ and $150$, the cluster remains aggregated but no longer rotates as a single coherent body.\\
These results show that active-mediated aggregation is robust in the larger system, whereas persistent collective rotation is favored for compact clusters that can sustain a coherent net torque.
\subsection{Role of chirality distribution}
\begin{figure}[H]
\centering
\includegraphics[width=7.2cm,height=6cm]{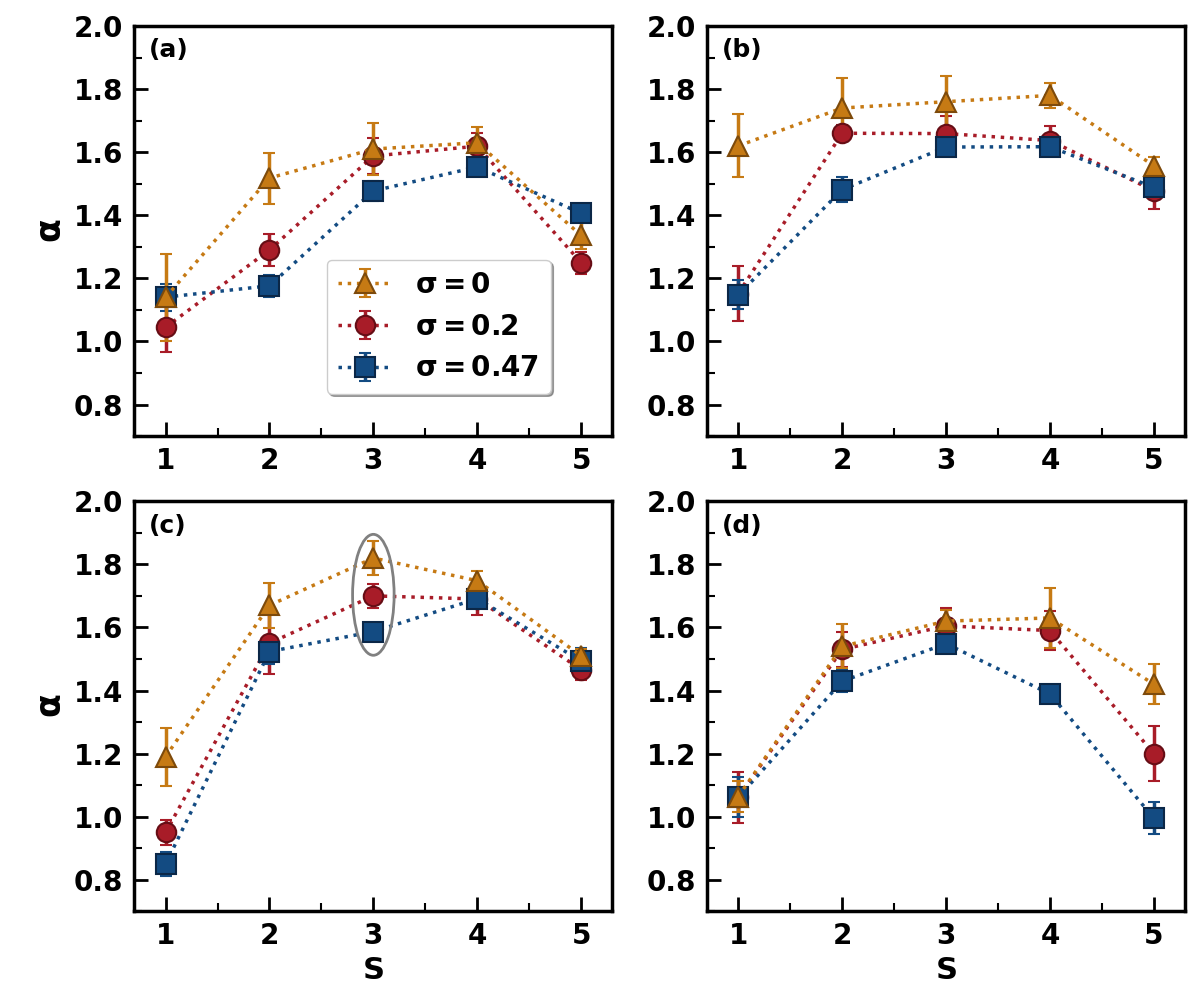}
\caption{Exponent $\alpha$ of the MSAD plotted as a function of the size ratio $S$. Data are shown for various standard deviations $(\sigma)$ of the chirality distribution, as labeled in each panel. The panels correspond to fixed active particle packing fractions of (a) $\phi_a=0.2$, (b) $\phi_a=0.3$, (c) $\phi_a=0.4$, and (d) $\phi_a=0.5$. Parameters enclosed within the ellipsoids correspond to the movies provided in Appendix \ref{sec:anim}. (Multimedia available online; see Movie $6$, $7$, and $8$)}
\label{fig9}
\end{figure}
We now investigate the role of intrinsic chirality fluctuations within the active bath. Fig.~\ref{fig9} (Multimedia available online) illustrates the MSAD exponent $\alpha$ as a function of the size ratio $S$ for varying standard deviations ($\sigma$) of a log-normal chirality distribution (discussed in Sec.~\ref{secII}). Specifically, we compare the initially considered distribution ($\sigma = 0.47$) against a narrower distribution ($\sigma=0.2$) and a uniform, constant-chirality bath ($\sigma=0$ and $\Omega_0=0.11$). For a clearer visualization of these dynamics, movies corresponding to the parameters inside the ellipsoidal data points of Fig.~\ref{fig9} (Multimedia available online) are provided in Appendix~\ref{sec:anim} (Movies $6,7$ and $8$).\\
For intermediate size ratios ($S=2,3,$ and $4$), the rotational dynamics depend strongly on  $\phi_a$. At intermediate packing fractions ($\phi_a=0.3$ and $0.4$), the system achieves its maximum rotational persistence. Specifically, for a constant-chirality bath ($\sigma = 0$), the chiral particles coordinate to exert a coherent, long-lived net torque, driving strongly superdiffusive or nearly ballistic angular motion, as shown in Fig.~\ref{fig9} (Multimedia available online). As the variance increases to $\sigma = 0.2$ and $0.47$ for these same intermediate $\phi_a$ ($0.3$ and $0.4$), local geometric frustration emerges, degrading the temporal coherence of the forcing and systematically lowering $\alpha$.\label{subsec:cd}
In contrast, at the extreme packing fractions $\phi_a$ ($0.2$ and $0.5$), the motion for these intermediate size ratios ($S = 2, 3,$ and $4$) remains superdiffusive but is dampened compared to the intermediate $\phi_a$  ($0.3$ and $0.4$). At $\phi_a=0.2$ for $S=2,3,$ and $4$, the active interactions are comparatively weak, yielding lower $\alpha$ values across all $\Omega$ distributions ($\sigma=0,0.2,$ and $0.47$) due to insufficient collective pushing. Conversely, at $\phi_a = 0.5$ for $S = 2, 3,$ and $4$, severe steric hindrance restricts rotational mobility compared to intermediate $\phi_a$ ($0.3$ and $0.4$).\\
For the smallest size ratio ($S = 1$), the rotational dynamics depend sensitively on both $\phi_a$ and the chirality variance ($\sigma$) of the bath. At extreme $\phi_a$ ($0.2$ and $0.5$), continuous structural disturbances from active particles comparable in size to the passive particles prevent the cluster from establishing a stable rotational axis, causing $\alpha$ to fall to the diffusive limit ($\alpha \approx 1.0$) across all distributions ($\sigma = 0, 0.2,$ and $0.47$). However, at intermediate $\phi_a$ ($0.3$ and $0.4$), a uniform constant chirality bath ($\sigma = 0$) imparts a sufficiently consistent net torque to drive strong superdiffusive motion. Introducing a distributed chirality ($\sigma = 0.2$ and $0.47$) at $S = 1$ for these intermediate $\phi_a$ ($0.3$ and $0.4$) significantly degrades the rotational coherence, pushing $\alpha$ to much lower values.\\
Finally, at the largest size ratio ($S = 5$), the exponent $\alpha$ drops across all  $\phi_a$, but for $\phi_a = 0.2,0.3$ and $0.4$, it remains nearly the same for $\sigma = 0,0.2$ and $0.47$, indicating that the large cluster size itself primarily limits the rotational persistence in this regime. For the highest packing fraction ($\phi_a =0.5$), the constant-chirality case ($\sigma = 0$) shows a slightly larger $\alpha$ than the distributed-chirality cases ($\sigma = 0.2$ and $0.47$). This suggests that, under strong crowding, a uniform chiral bath can still maintain somewhat better rotational coherence (see Appendix~\ref{sec:cdis}), whereas heterogeneity in chirality further weakens the collective forcing and shifts the dynamics closer to the diffusive regime. 
\section{Discussion \label{secIV}}
The results establish a clear phenomenology for passive clusters immersed in a chiral active bath. Persistent rotation is not a generic consequence of activity or chirality alone; instead, it appears only in a restricted region of parameter space, namely at intermediate size ratios and moderate active packing fractions. Outside this window, the clusters remain predominantly diffusive or rotate only intermittently, showing that coherent rotation requires a specific combination of geometry, forcing, and structural stability.\\
The structural analysis explains why this window is so narrow. In the rotating regime, the passive clusters retain local ordering, as reflected in the higher-order peaks of $g_{pp}(r_s)$, whereas stronger crowding or extreme size ratios disrupt that order and promote internal rearrangements or breakup. Thus, rotation occurs only when the cluster is ordered enough to behave collectively but not so rigid that it loses responsiveness to the active bath.\\
The asphericity $A_p$ and the mean torque $T_p$ further clarify this balance. Intermediate size ratios produce elongated clusters that can couple efficiently to the active forcing, and this geometric anisotropy is accompanied by a strong net torque. However, the largest torque does not automatically produce rigid-body rotation: for the largest clusters, the same forcing is dissipated through internal rearrangements rather than coherent motion, so torque must be understood together with cohesion and shape.\\
The dynamical signatures are consistent with this interpretation. The angular autocorrelation functions $(C_{\theta}(\tau))$ exhibit long-lived oscillations only in the same parameter window where asphericity and torque are enhanced, and the mean-squared angular displacement becomes superdiffusive there as well. Together, these observables show that the cluster acquires a genuine rotational memory rather than a sequence of transient angular kicks.\\
Chirality heterogeneity provides an additional control parameter. A uniform chirality distribution supports the most coherent rotation, whereas increasing chirality fluctuations progressively suppress angular persistence and drive the system toward diffusion. This indicates that coherent rotation requires not only geometric compatibility but also sufficiently synchronized orientational forcing from the active bath.\\
Translational motion is less selective than angular motion. The cluster center of mass shows the usual ballistic-to-diffusive crossover across the full parameter range, but the long-time diffusivity is not maximized in the same region as the rotational response. This separation between translational and rotational optimization shows that the passive cluster behaves like an emergent active object, yet translation and rotation are controlled by different structural constraints.\\
The system-size study reinforces this distinction. Aggregation remains robust in the larger domain, but coherent rotation weakens as the passive cluster grows because the active bath can no longer impose a uniform torque over the full structure. In that sense, large systems preserve clustering but lose the compactness needed for rigid-like collective rotation.\\
Taken together, these results demonstrate that the emergence of persistent cluster rotation in chiral active baths is a highly collective effect. It appears only when structural coherence, geometric anisotropy, strong net torque, and a homogeneous active bath coexist, emphasizing the complex interplay between geometry and local orientational correlations in active-passive mixtures.\\
\begin{figure*}[!hbtp]
\centering
\includegraphics[width=13cm, height=8cm]{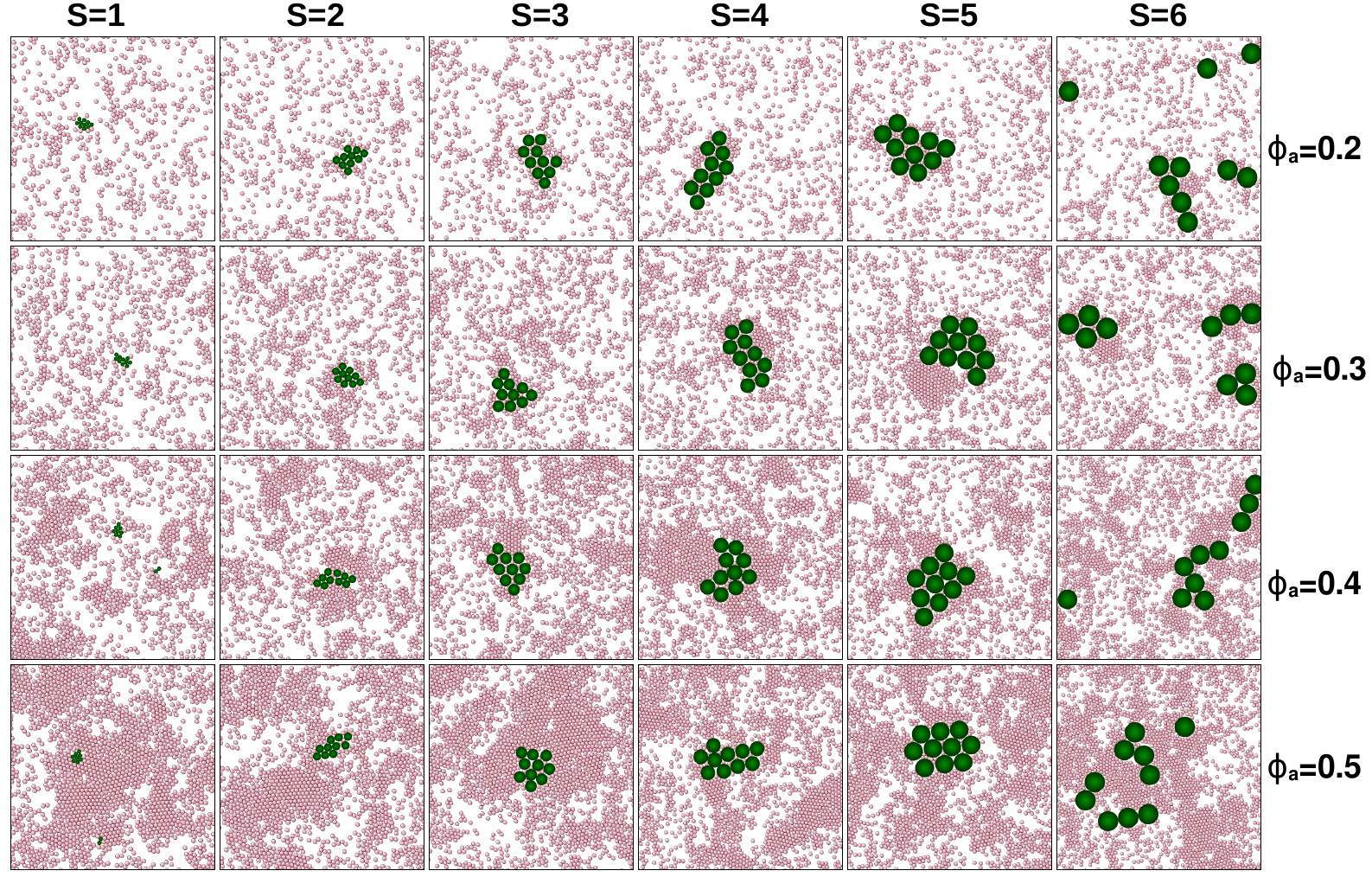}
\caption{Snapshots of the active-passive mixture at time $t = 10^3\tau_p$. Columns correspond to different size ratios $S=1,2,3,4,5,$ and $6$ as labeled at the top. Rows correspond to different values of the active particle area fraction $\phi_a = 0.2, 0.3, 0.4,$ and $0.5$, as labeled on the right. (Multimedia available online)}
\label{fig10}
\end{figure*}
The present results are most directly applicable to effectively overdamped systems in which substrate friction dominates, such as active colloids confined near surfaces, vibrated granular active matter, and related synthetic active-particle assemblies. This interpretation is consistent with our previous combined experimental and numerical study of passive colloids in active liquids \cite{kushwaha2023phase}, in which bacteria-driven colloids exhibited dynamic clustering, and dry active-particle simulations reproduced the essential size-ratio dependence of active-mediated assembly. These observations suggest that dry active-matter models capture the fundamental mechanisms underlying active-mediated aggregation and collective organization.\\
In wet active systems, however, fluid-mediated interactions can modify the effective forces, torques, and transport of passive clusters, and may therefore shift the quantitative phase boundaries, cluster stability, and rotational persistence reported here. Extending the present framework to explicitly incorporate hydrodynamic interactions therefore represents an important direction for future work.\\
Understanding this spontaneous conversion of nonequilibrium energy into coherent rotation may help guide the design of self-guided microrobots and self-assembling micro-gears capable of extracting mechanical work from active environments \cite{barona2024playing,dhatt2023accelerating}.

\section{Data availability}	
The data that support the findings of this study are available within the article.
\section{Acknowledgments}
The authors gratefully acknowledge the support and resources provided by the PARAM Shivay Facility under the National Supercomputing Mission, Government of India, at the Indian Institute of Technology (BHU), Varanasi. We also thank Vijay Chikkadi and Pragya Kushwaha for their insightful discussions. S.M. acknowledges financial support from DST-SERB India (Grant Nos. ECR$/2017/000659$, CRG$/2021/006945$, and MTR$/2021/000438$) and D.K. acknowledges financial support from UGC India.sh
\section{Appendix A: snapshots}\label{sec:app}
Fig.~\ref{fig10} (Multimedia available online) shows representative steady-state snapshots of the chiral active–passive mixture for size ratios $S=1,2,3,4,5,6$ and active packing fractions $\phi_a=0.2,0.3,0.4,0.5$. These snapshots provide a qualitative view of how chiral active particles distribute around passive aggregates and support the trends in clustering and internal ordering discussed in the main text.
\section{Appendix B: Passive-Passive interaction}\label{sec:force}
\begin{figure}[H]
\centering
\includegraphics[width=8cm, height=5cm]{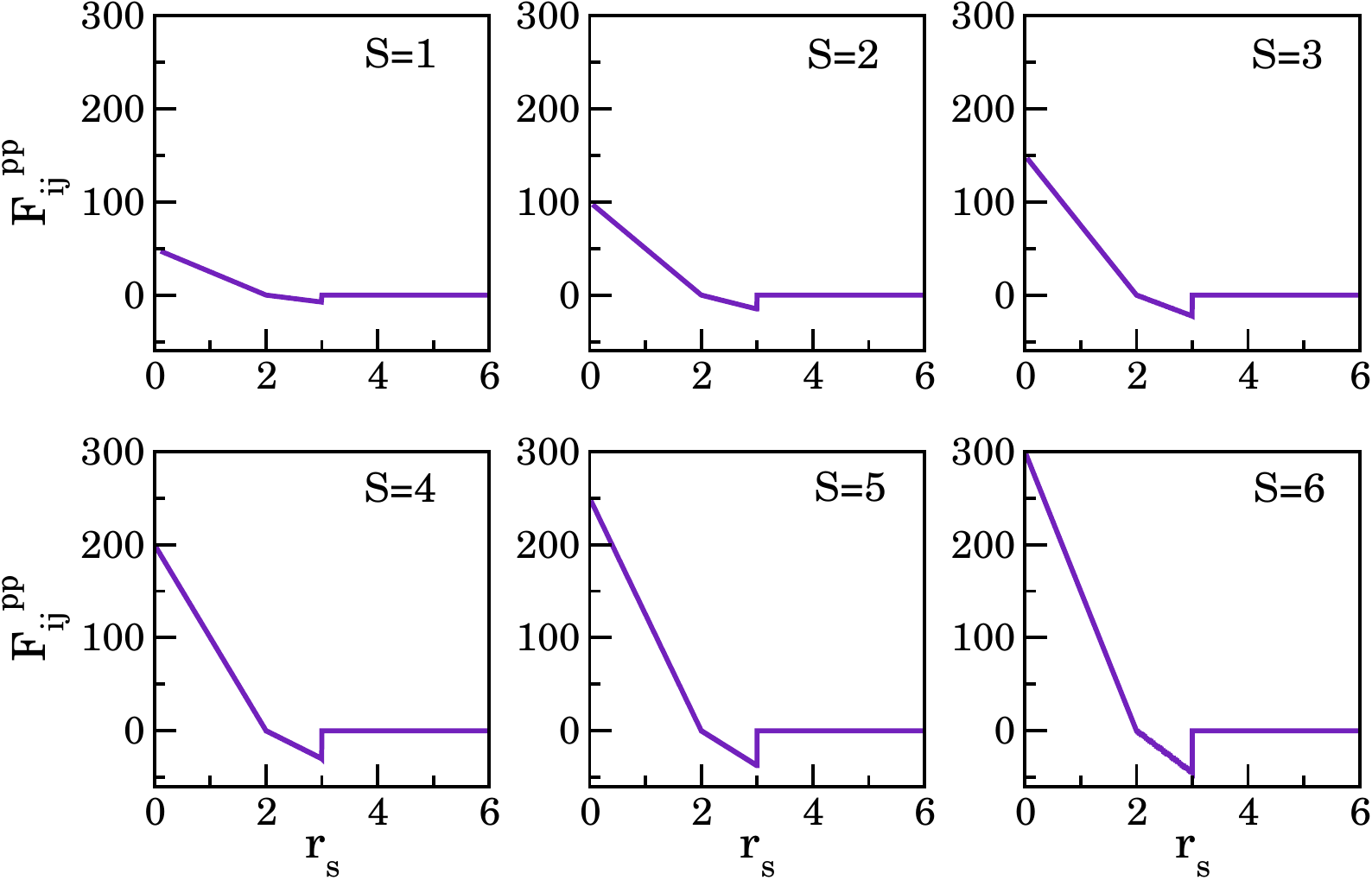}
\caption{Passive-passive interaction force $F_{ij}^{pp}$ vs. scaled separation $r_s$, similar to Fig.~\ref{fig2}, for different size ratios $S$ as labeled in the plot. (Multimedia available online)}
\label{fig11}
\end{figure}

Fig.~\ref{fig11} (Multimedia available online) illustrates the passive-passive interaction force profile $F^{pp}_{ij}$ as a function of scaled separation $r_s$ for various size ratios $S$. The force combines steep, short-range repulsion due to volume exclusion with a weak attractive tail that mimics cohesive interactions between the passive particles. As shown by the analytical form of the force, increasing the size ratio $S$ linearly increases both the maximum magnitude of the attractive force and its spatial range, which strengthens the internal cohesion of larger passive clusters against the active bath.\\
\begin{figure*}[!hbtp]
\centering
\includegraphics[width=18cm, height=12cm]{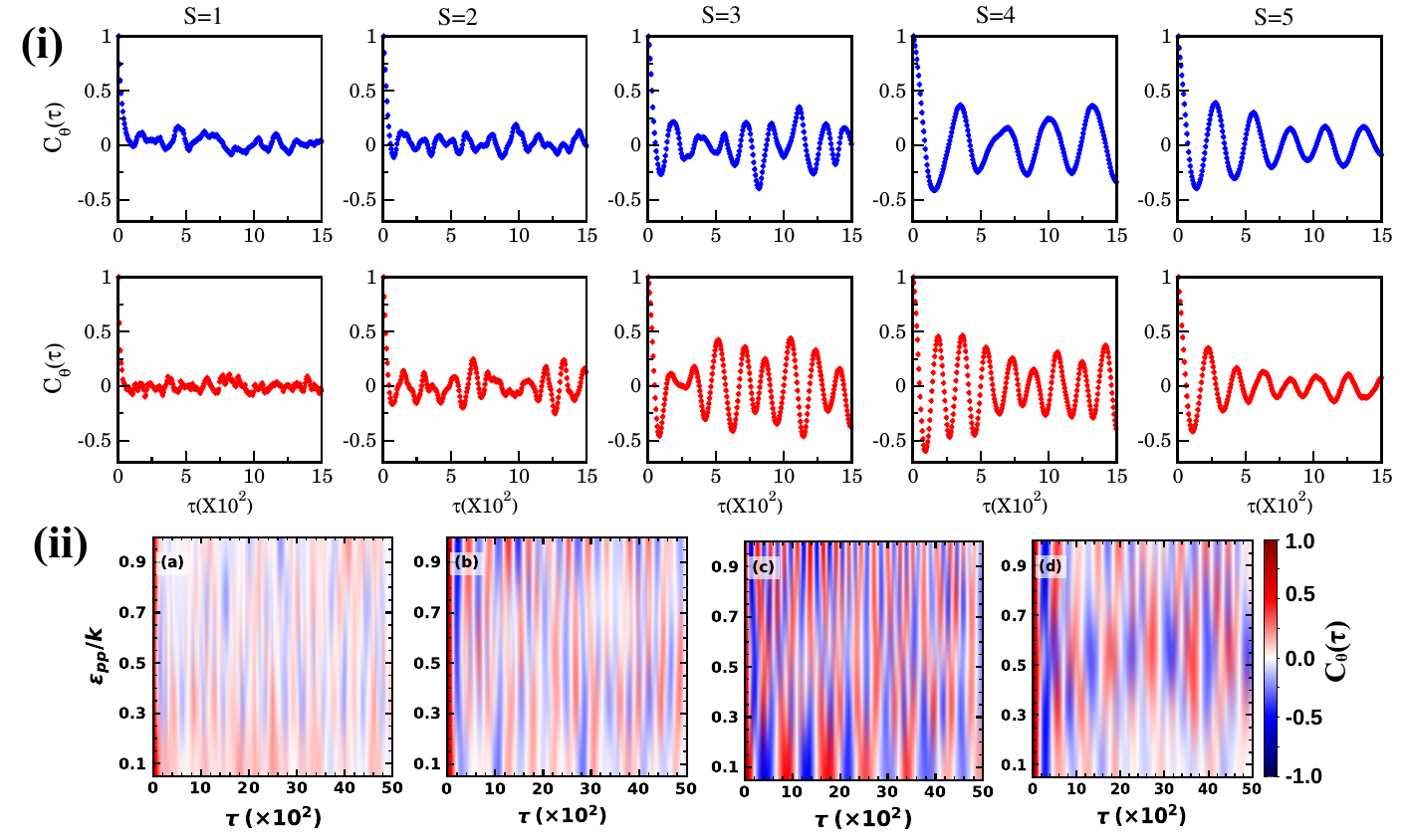}
\caption{(i) Angular autocorrelation function $C_\theta(\tau)$ as a function of the lag time $\tau$ (in units of $\tau_p$) for different size ratios $S$, as labeled in the panels, at a fixed active-particle area fraction $\phi_a=0.4$. The blue curves (first row) correspond to the attraction strength $\epsilon_{pp}=0.1k$, while the red curves (second row) correspond to $\epsilon_{pp}=0.6k$. (ii) Heat-map representation of the angular autocorrelation function $C_\theta(\tau)$ as a function of the lag time $\tau$ and the normalized attraction strength $\epsilon_{pp}/k$ for (a) $S=2$, (b) $S=3$, (c) $S=4$, and (d) $S=5$ at $\phi_a=0.4$. Red and blue regions denote positive and negative angular correlations, respectively. (Multimedia available online; see Movies $9$ and $10$.)}
\label{fig12}
\end{figure*}
To further examine the sensitivity of the rotational state to the passive-passive attraction strength, Fig.~\ref{fig12}(ii) (Multimedia available online) presents heat maps of $C_\theta(\tau)$ as a function of lag time $\tau$ and attraction strength $k$ at fixed $\phi_a=0.4$. Persistent oscillatory bands are observed over a broad range of attraction strengths for all investigated size ratios, demonstrating that cluster rotation is not confined to a narrow attraction window.
In particular, for $S=3$ and $S=4$, clear oscillatory correlations are already present at the lowest attraction strength ($\epsilon_{pp}=0.1k$), indicating that persistent rotation survives even when the passive cluster is weakly bound. As the attraction strength increases, the alternating red and blue bands become more closely spaced, reflecting an increase in the rotational frequency. The strongest and most coherent oscillatory patterns occur for intermediate size ratios, especially $S=4$, where collective rotation remains robust across the entire range of attraction strengths considered.\\
For clearer visualization of this behavior, Fig.~\ref{fig12}(i) (Multimedia available online) shows the angular autocorrelation function $(C_\theta(\tau))$ for two representative attraction strengths, $(\epsilon_{pp}=0.1k)$ and $(\epsilon_{pp}=0.6k)$. These cases may be compared with the baseline attraction strength $(\epsilon_{pp}=0.3k)$ used throughout the main text, whose corresponding $(C_\theta(\tau))$ is shown in the third row $(\phi_a=0.4)$ of Fig.~\ref{fig4} (Multimedia available online). The associated cluster dynamics are presented in Movies $9$ and $10$, providing a direct visualization of the rotational behavior at weak and strong attraction strengths.
\section{Appendix C: NEIGHBOR LIST UPDATE}\label{sec:nbd}
\begin{figure}[H]
\centering
\includegraphics[width=8cm, height=6.5cm]{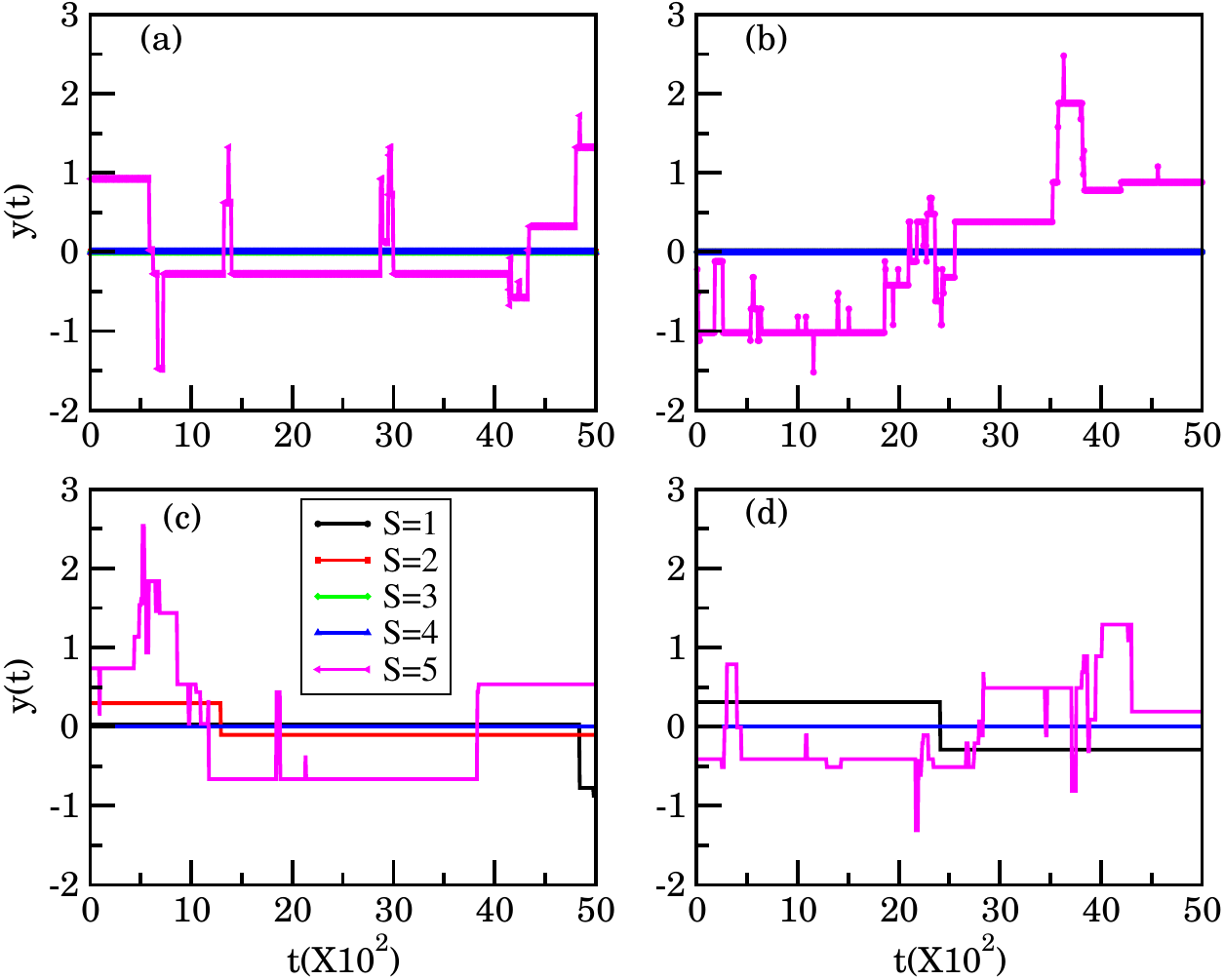}
\caption{Variation of the update in the neighbor list $y(t)$ of passive particles in the cluster as a function of time $t$ (in units of $\tau_p$). Data are shown for various size ratios $S$ at fixed active packing fractions: (a) $\phi_a=0.2$, (b) $\phi_a=0.3$, (c) $\phi_a=0.4$, and (d) $\phi_a=0.5$. (Multimedia available online; see Movie $1$ and $2$)}
\label{fig13}
\end{figure}

To characterize how the local neighborhood changes within the cluster, we track updates in the neighbor list \cite{pattanayak2020speed,singh2021bond} of the passive particles using the observable $y(t) = \left\langle \left( N_R^i(t) \times \frac{N_p}{2} \right) - \sum_{j \in R } j \right\rangle \cdot N_R^i$. Here, $N_R^i$ is the number of passive particles within the interaction radius $R = 2a_2 + S a_1$ of the $i^{\text{th}}$ passive particle, which corresponds to the range of attraction. $N_p$ is the total number of passive particles, and the summation runs over the fixed permanent indices $j$ of all passive particles currently inside that interaction radius. The notation $\langle \cdots \rangle$ denotes averaging over all passive particles in the system.\\
Physically, this expression works by tracking the specific identities of neighboring particles over time. If the cluster behaves as a solid rigid body, each particle maintains the same set of neighbors. As a result, the sum of their indices remains constant, and $y(t)$ exhibits a flat trajectory. Conversely, when particles shuffle and exchange positions, the specific neighbor indices $j$ within the interaction radius change, causing the value of $y(t)$ to fluctuate. Therefore, temporal fluctuations in $y(t)$ serve as a direct mathematical signature of internal structural rearrangements.\\
The time series of $y(t)$ remains stable with minimal fluctuations for most values of $S$, indicating that these smaller clusters maintain their structural integrity during motion. In contrast, at $S = 5$ across all $\phi_a$, $y(t)$ shows pronounced and continuous fluctuations around zero (Fig.~\ref{fig13} (Multimedia available online)). The high frequency of these fluctuations confirms that the neighbor lists are constantly updated by internal particle shuffling. Furthermore, variations in the overall magnitude of $y(t)$ imply dynamic changes in the total number of neighbors within the interaction radius of a passive particle. This distinct structural yielding at $S=5$, compared to the rigid-body behavior of smaller clusters, is directly corroborated by the visual evidence in Movie $1$ and Movie $2$ (Appendix~\ref{sec:anim}).
\section{Appendix D: Varying the Chirality Distribution}\label{sec:cdis} 
\begin{figure}[H]
\centering
\includegraphics[width=8cm, height=10cm]{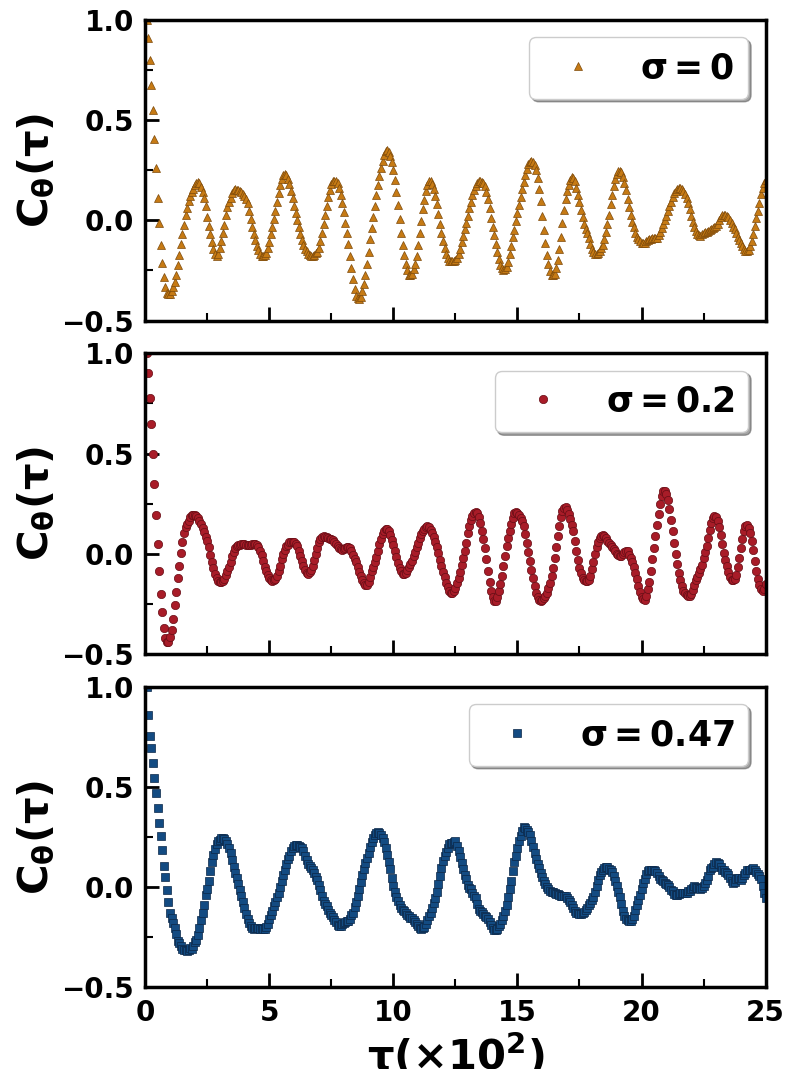}
\caption{Angular autocorrelation function $C_\theta(\tau)$ plotted as a function of the time lag $\tau$ (in units of $\tau_p$). The data are shown for a fixed size ratio $S=3$ and packing fraction $\phi_a=0.4$ across different chirality distribution variances $\sigma=0, 0.2,$ and $0.47$. (Multimedia available online; see Movie $6$, $7$, and $8$)}
\label{fig14}
\end{figure}

Fig.~\ref{fig14} (Multimedia available online) investigates the effect of chiral active bath heterogeneity by plotting the angular autocorrelation function $C_\theta(\tau)$ for the specific parameters highlighted by the ellipsoid in Fig.~\ref{fig9} (Multimedia available online) ($S=3$ and $\phi_a=0.4$) across different chirality distribution variances ($\sigma$). When the active bath maintains a uniform, constant chirality ($\sigma=0$), the chiral particles coordinate to exert a highly coherent, long-lived net torque on the passive cluster. In the plot, this drives robust rotation characterized by rapid, clear periodic oscillations with deep negative dips, indicating that the cluster completes full revolutions before its orientational memory decorrelates.\\ 
However, introducing heterogeneity into the chirality distribution ($\sigma=0.2$ and $\sigma=0.47$) degrades the orientational coherence of the active forcing. As a result, the plotted angular autocorrelation decays more quickly, and the periodic oscillations visibly dampen and stretch over longer time lags. This illustrates the systematic dampening effect of chirality variance on the cluster's rotational persistence, shifting the dynamics closer to a diffusive state as the variance increases. These corresponding dynamical behaviors are explicitly visualized in the supplementary animations, where Movie $6$ shows robust collective rotation at $\sigma=0$, Movie $7$ illustrates dampened rotation at $\sigma=0.2$, and Movie $8$ demonstrates degraded rotational coherence at $\sigma=0.47$.
\section{Appendix E:Movies}\label{sec:anim} 
 \textbf{Neighbour List Update:} In Movies $1$ and $2$, we show the updated neighbor list. In Movie $1$, the cluster rotates steadily with stable particle identities in local neighborhoods. Movie $2$ reveals active particle exchanges within the cluster of passive particles, with particles visibly trading neighbors, as discussed in Appendix~\ref{sec:nbd}.\\
\textbf{Movie 1}: Shows a passive cluster rotation for the size ratio $S=4$ and the active packing fraction $\phi_a=0.3$.\\ 
\textit{Link:} \textcolor{blue}{\url{https://drive.google.com/file/d/1ZEzO4rdhE-kdS0LN1ExmoEcEqJSqd8Hq/view?usp=drive_link}}.\\
\textbf{Movie 2}: Shows a passive cluster rotation for the size ratio $S=5$ and the active packing fraction $\phi_a=0.3$.\\
\textit{Link:} \textcolor{blue}{\url{https://drive.google.com/file/d/1dk7SMm5b0EOaOcuH0v2QKQ7_bL8TQvFH/view?usp=drive_link}}.\\
\textbf{Angular Autocorrelation $C_\theta(\tau)$:}
In Movies $3,4$ and $5$, we have shown the animation of a rotating cluster with autocorrelation $C_\theta(\tau)$. In the plot, the color is as shown in Fig.~\ref{fig1}(d) (Multimedia available online), representing time.\\
\textbf{Movie 3}: Shows a passive cluster rotation with $C_\theta(\tau)$ for the size ratio $S=2$ and  the active packing fraction $\phi_a=0.5$.\\
\textit{Link:} \textcolor{blue}{\url{https://drive.google.com/file/d/1NKjCh9H7GhdAqwnm-64suM3GxG2QDZZy/view?usp=drive_link}}\\
\textbf{Movie 4}: Shows a passive cluster rotation with $C_\theta(\tau)$ for the size ratio $S=4$ and the active packing fraction $\phi_a=0.3$.\\ \textit{Link:} \textcolor{blue}{\url{https://drive.google.com/file/d/1LtaPkwa7AsQr68P4YmGVz7hQYqAiw8v3/view?usp=drive_link}}.\\
\textbf{Movie 5}: Shows a passive cluster rotation with $C_\theta(\tau)$ for the size ratio $S=5$ and  the active packing fraction $\phi_a=0.5$.\\
\textit{Link:} \textcolor{blue}{\url{https://drive.google.com/file/d/1jRrp2qDvP07XeHZskIXu2NLg5nYua_-T/view?usp=drive_link}}\\
\textbf{Role of chirality distribution:} In Movies $6,7$ and $8$, we show the rotation of passive clusters immersed in a chiral active bath for $\phi_a=0.4$ and $S=3$, with one passive particle highlighted in red to 
 clearly visualize the complete rotation cycle of the cluster.\\
\textbf{Movie 6}: Shows passive cluster rotation for an active-particle chirality distribution with $\sigma=0$. The cluster completes an oscillation in $287.5\tau_p$.\\
\textit{Link:} \textcolor{blue}{\url{https://drive.google.com/file/d/1PcYQCr_uTktZqTDoAHDpbodP0ZJ5pCMC/view?usp=drive_link}}\\
\textbf{Movie 7}: Shows passive cluster rotation for an active-particle chirality distribution with $\sigma=0.2$. The cluster completes an oscillation in  $312.5\tau_p$.\\ 
\textit{Link:} \textcolor{blue}{\url{https://drive.google.com/file/d/1IB2dTlVZJJZthpmOiFnIG_quszWgePiR/view?usp=drive_link}}.\\
\textbf{Movie 8}: Shows passive cluster rotation for an active-particle chirality distribution with $\sigma=0.47$. The cluster completes an oscillation in $350.0\tau_p$.\\
\textit{Link:} \textcolor{blue}{\url{https://drive.google.com/file/d/14D3KH2vofKQt_oQYST96OuoNFp123yPl/view?usp=drive_link}}.\\
\textbf{Passive-Passive Interaction:} In Movies $9$ and $10$, we observe the effect of passive-passive attraction strength.  \\
\textbf{Movie 9}: Shows passive cluster rotation in a chiral active bath with a passive-passive attraction strength of $\epsilon_{pp}=0.1k$.\\
\textit{Link:} \textcolor{blue}{\url{https://drive.google.com/file/d/1F1i0BmUabeBPKB9WUHM8aj0rXYtdw5vV/view?usp=drive_link}}.\\
\textbf{Movie 10}: Shows passive cluster rotation in a chiral active bath with a passive-passive attraction strength of $\epsilon_{pp}=0.6k$.\\
\textit{Link:} \textcolor{blue}{\url{https://drive.google.com/file/d/1fCyY8efVHlwzeHhIjvtj74haWEBJ-VY8/view?usp=drive_link}}\\
\textbf{Role of system size and packing fraction of the passive particles:}
In Movies $11$, $12$, and $13$, we demonstrate the effect of increasing the passive cluster size in a larger simulation box ($L=200a_1$) on the rotational dynamics of the passive aggregate.\\
\textbf{Movie 11}:Shows the dynamics of the passive cluster for $N_p=10$ in a larger system ($L=200a_1$), illustrating persistent collective rotation of the passive cluster.\\
\textit{Link:} \textcolor{blue}{\url{https://drive.google.com/file/d/17f1LJtm7oUNewvn-innb14UxmdJsQBek/view?usp=sharing}}.\\
\textbf{Movie 12}: Shows the dynamics for $N_p=30$, where the passive cluster remains compact and exhibits persistent collective rotation.\\
\textit{Link:} \textcolor{blue}{\url{https://drive.google.com/file/d/1EpqHY9D0wc-yF60YxY-4vE4vh6p1SgrV/view?usp=share_link}}.\\
\textbf{Movie 13}: Shows the dynamics for $N_p=100$, where the passive particles remain aggregated, but the large cluster no longer exhibits coherent, persistent rotation.\\
\textit{Link:} \textcolor{blue}{\url{https://drive.google.com/file/d/1XzRTjbpg5iXYbcZJie2D_2MPOXFEQDhY/view?usp=share_link}}.\\
\bibliography{citation}
\end{document}